\documentclass[12pt]{article}

\usepackage{setspace}

\usepackage[utf8]{inputenc} 
\usepackage[T1]{fontenc}    

\usepackage{hyperref}       
\usepackage{url}            
\usepackage{booktabs}       
\usepackage{amsfonts}       
\usepackage{nicefrac}       
\usepackage{microtype}      
\usepackage{lipsum}
\usepackage{fancyhdr}       
\usepackage{graphicx}       
\graphicspath{{graphs/}}     
\usepackage{caption}
\usepackage{subcaption}
\usepackage[dvipsnames]{xcolor}
\usepackage{cite}
\usepackage{epsfig}
\usepackage{textcomp}
\usepackage{bm}

\usepackage{algorithm}
\usepackage{algorithmic}
\usepackage{dsfont}
\usepackage{mathrsfs}
\usepackage{multirow, makecell}
\usepackage{tabularray}
\usepackage{soul}

\usepackage{pdflscape}
\usepackage{amsmath}
\usepackage{colortbl}

\begin{document}

\newpage
\thispagestyle{empty}


\newpage
\clearpage

\pagestyle{fancy}
\thispagestyle{empty}
\rhead{ \textit{ }} 
\setlength{\tabcolsep}{15pt}

\fancyhead[LO]{Compound V3 Economic Audit Report}

\title{Compound V3 Economic Audit Report}
\author{
  by \\
 Rik Ghosh, Samrat Gupta, \\ Arka Datta, Abhimanyu Nag, Sudipan Sinha \footnote{Correspondence : sudipan@chainrisk.cloud}\\
  \textbf{Chainrisk}}

\date{August 2024}

\maketitle

\begin{abstract}
Compound Finance is a decentralized lending protocol that enables the secure and efficient borrowing and lending of cryptocurrencies, utilizing smart contracts and dynamic interest rates based on supply and demand to facilitate transactions. The protocol enables users to supply different crypto assets and accrue interest, while borrowers can avail themselves of loans secured by collateralized assets. Our collaboration with Compound Finance focuses on harnessing the power of the Chainrisk simulation engine to optimize risk parameters of the Compound V3 (Comet) protocol. This report delineates a comprehensive methodology aimed at calculating key risk metrics of the protocol. This optimization framework is pivotal for mitigating systemic risks and enhancing the overall stability of the protocol. By leveraging Chainrisk's Cloud Platform, we conduct millions of simulations to evaluate the protocol's Value at Risk (VaR) and Liquidations at Risk (LaR), ultimately providing recommendations for parameter adjustments.

\vspace{0.5cm}
\end{abstract}

\section{Disclaimer}	
This report is presented for informational purposes only and does not constitute an offer by Chainrisk to sell, purchase, or recommend any securities. It is not intended to solicit financial advice services or other related services. The report does not promote or endorse any specific security purchase, sale, retention, or reference. It should not be interpreted as an invitation to engage in financial transactions or be considered investment advice or services for any individual in any jurisdiction. The views expressed herein are not recommendations to buy, sell, or hold any securities, nor does this report offer any assessment of the suitability of any investment product. Readers are cautioned against relying solely on the information presented to make investment decisions. 

The information has not been tailored to any specific investor's needs, objectives, or financial situation. The investments discussed may not be suitable for all investors, and the information is provided without consideration for any recipient's unique needs, financial status, or investment goals.

Finally, any views expressed in this report are based on the information available during writing and may change due to new or updated information. All information presented is subject to correction and may become unreliable due to changes in market conditions or economic circumstances.

\section{About Chainrisk}	
Chainrisk is a Security SaaS company building a unified simulation platform that allows teams to test protocols efficiently while understanding how they will react to adversarial market environments. The backbone of our technology is a cloud-based, agent- and scenario-based simulation engine that allows users to create specific market scenarios to test new features and assets to understand risk parameters better.

\section{Background}
\subsection{Overview}
Compound Finance is a decentralized lending protocol that facilitates the borrowing and lending of cryptocurrencies. Compound has evolved through various iterations, with Compound V3 (Comet) being its latest version. This iteration introduces a more capital-efficient market structure, allowing for single-asset borrowing while optimizing liquidity and enhancing user experience. The architecture of Compound V3 is designed to streamline the lending process, incorporating advanced smart contract functionalities that enable automatic interest rate adjustments based on supply and demand dynamics.

\subsection{Scope of Audit}
The scope of the audit is the native USDC market on Arbitrum One. We will conduct a comprehensive examination of the collateral assets in the USDC market, specifically Wrapped Ether (WETH), Wrapped Bitcoin (WBTC), GMX, and Arbitrum (ARB). It is to be noted that for the rest of the report, Ethereum (ETH) will be interchangeably used with Wrapped Ether (WETH), and is also the same for Bitcoin (BTC) and Wrapped Bitcoin (WBTC). We will focus on stress testing the protocol and recommending risk parameters on the Arbitrum USDC market. 

\subsection{Goals of the Analysis}
In this paper, we provide the market risk calculation framework that Chainrisk Labs will utilize to determine risk parameters for the Compound V3 protocol. We specify our methodology for finding optimal risk parameters that juggle capital efficiency and protocol risk, and we use Chainrisk’s agent-based simulation platform to estimate protocol losses. This paper is a technical deep dive into the process that we will use to recommend Compound protocol parameters.

The primary objective of this analysis is to stress test the Compound V3 market under adverse conditions to evaluate its resilience and performance metrics. Key metrics such as Value at Risk (VaR) and Liquidations at Risk (LaR) within the Arbitrum USDC market will be calculated to assess the protocol's risk exposure. Based on the findings, we will provide tailored parameter recommendations aimed at optimizing the Arbitrum USDC market on the Arbitrum chain, thereby enhancing its stability and mitigating potential systemic risks.

\section{Protocol Overview}
\subsection{Compound V3 Overview}
Compound V3, often referred to as Comet, is a significant evolution of the Compound protocol designed to optimize decentralized lending and borrowing. Unlike previous versions, In Compound V3, each market is defined by a single base asset, which serves as the borrowable asset, and allows for the use of multiple collateral assets within that market. Compound V3's focus on a single base asset represents a strategic departure from traditional decentralized lending platforms. This design enhances security and capital efficiency by isolating borrowing and collateralization processes.
The protocol leverages a token-based incentive structure, distributing COMP tokens to liquidity providers to attract capital and foster protocol growth. Compound V3 employs a token-based incentive structure to attract liquidity providers and borrowers. By supplying assets to the protocol, users earn COMP tokens as a reward. This mechanism is pivotal for maintaining liquidity within the ecosystem, ensuring that users are rewarded for their contributions to the overall health of the platform.

\subsubsection{Collateral and Borrowing}
In Compound V3, borrowing is an over-collateralized process where users must supply collateral assets to borrow the base asset. This over-collateralization serves as a safeguard against market volatility and potential insolvency, ensuring that the protocol maintains a robust collateralization ratio.
For example, consider a scenario where a user intends to borrow USDC by using ETH as collateral. \\
Collateral Asset: ETH \\
Base Asset: USDC \\
Borrow Collateral Factor for ETH: $75\%$ \\ 
User Supplies: $3$ ETH, valued at $\$9,000$ (assuming an ETH price of $\$3,000$). 

Calculation of Borrowing Capacity:
The borrowed collateral factor indicates the percentage of the collateral that can be borrowed. In this case, the user can borrow up to $75\%$ of the value of their supplied ETH. This factor signifies that for every dollar of ETH supplied, the platform permits borrowing up to $\$0.75$ of USDC.

To calculate the borrowing capacity:
Borrowing Capacity $=$ Value of collateral Supplied $\times$ Collateral Factor
i.e. Borrowing Capacity $= \$9000 \times 0.75 = \$6750$. 
This means the user can borrow a maximum of $\$6,750$ worth of USDC against their ETH collateral.

\subsubsection{Interest Rate Curve}
Compound V3 employs a dynamic interest rate model determined by asset utilization. Users supplying assets to a market earn interest based on a supply rate curve, while borrowers pay interest based on a separate borrow rate curve. The interest rate model features a utilization rate "kink," a critical threshold beyond which the interest rate increases more steeply.

The interest rate for supplying assets is directly influenced by the utilization rate but operates independently of the borrowing interest rate, which follows a separate curve. It is important to note that collateral assets do not accrue interest or yield returns.

The formulas for calculating the supply and borrow rates are as follows:

\begin{itemize}
    \item \textbf{Supply Rate}:
        The supply rate is a function of the utilization rate and is calculated as follows:

            \begin{itemize}
                \item Below the Kink Point:
                
                Supply Rate $ = r_{base}^s + s_{low}^s \times u$, where, \\ $r_{base}^s =$ Supply per second interest rate base, \\ $s_{low}^s =$ Supply per second interest rate slope (below the kink), \\ $u = $utilization rate.
                \item Above the Kink Point:
                
                Supply Rate $ = r_{base}^s + s_{low}^s \times k^s + s_{high}^s \times (u-k^s)$, where, \\ $r_{base}^s =$ Supply per second interest rate base, \\ $s_{low}^s =$ Supply per second interest rate slope (below the kink), \\ $s_{high}^s =$ Supply per second interest rate slope (above the kink), \\ $k^s =$ Supply kink utilization rate, \\ $u = $utilization rate.
            \end{itemize}
    \item \textbf{Borrow Rate} :
            The borrowing rate is also a function of the utilization rate and follows a similar structure:
           \begin{itemize}
                \item Below the Kink Point:
                
                Borrow Rate $ = r_{base}^b + s_{low}^b \times u$, where, \\ $r_{base}^b =$ Borrow per second interest rate base, \\ $s_{low}^b =$ Borrow per second interest rate slope (below the kink), \\ $u = $utilization rate.
                \item Above the Kink Point:
                
                Borrow Rate $ = r_{base}^b + s_{low}^b \times k^b + s_{high}^b \times (u-k^b)$, where, \\ $r_{base}^b =$ Borrow per second interest rate base, \\ $s_{low}^b =$ Borrow per second interest rate slope (below the kink), \\ $s_{high}^b =$ Borrow per second interest rate slope (above the kink), \\ $k^b =$ Borrow kink utilization rate, \\ $u = $utilization rate.
            \end{itemize}
            
\end{itemize}

\subsubsection{Liquidation Mechanism}
The liquidation mechanism in Compound V3 is a critical component designed to maintain the protocol's solvency. Liquidation is determined by the liquidation collateral factor. A borrower's position is liquidated when the collateral value falls below a predefined liquidation collateral factor.

In Compound V3, The liquidation mechanism is designed to maintain the protocol's solvency by allowing the protocol to absorb collateral when a borrower's account becomes eligible for liquidation, effectively settling the outstanding debt 
The absorbed collateral is then put up for sale, with the sale price determined by the Liquidation Penalty (LP) and StoreFront Price Factor (SFP). Anyone can buy this discounted collateral asset and sell it on an exchange to profit on the difference. The collateral becomes available for purchase by liquidators only when the current reserve remains below the designated target reserve.

The sale price of the collateral is determined by two key factors: LP and SFP. The LP is the penalty incurred by borrowers during liquidation, which reduces the amount they receive. Meanwhile, the SFP sets the percentage of the LP allocated to buyers of the absorbed collateral, influencing the final sale price.

\textbf{To illustrate the liquidation process in Compound V3, consider the following example:}
\begin{itemize}
    \item Collateral: $1$ ETH, valued at $\$3,000$.
    \item Borrowed Amount: $\$2,400$ ($80\%$ of the collateral value).
    \item Borrow Collateral Factor (BCF): $80\%$.
    \item Liquidation Collateral Factor (LCF): $92\%$.
    \item Liquidation Factor (LF): $95\%$.
    \item Liquidation Penalty (LP): $5\%$, calculated as $1-LF$.
    \item StoreFront Price Factor (SFP): $60\%$.
\end{itemize}

Price Drop Scenario
Assume the price of ETH drops by $10\%$,  the new collateral value is calculated as follows:
New Collateral Value $= \$3,000 \times (1-0.10) = \$2,700$. 

At this point, the borrowed amount remains at $\$2,400$. Since the collateral value of $\$2,700$ is below the liquidation collateral factor of $\$2,760$ (calculated as Borrowed Amount×0.92), the account is now eligible for liquidation.

\textbf{Upon triggering liquidation, the following steps are executed:}

\textbf{Collateral Absorption:} The protocol absorbs the entire collateral of $1$ ETH.

\textbf{Liquidation Payment:} The borrower receives a payment based on the collateral value, adjusted for the LP:

\begin{align*}
    \text{Payment} &= \text{Collateral Value} \times (1 - \text{LP}) \\
    &= \$2700 \times (1 - 0.05) \\
    &= \$2700 \times 0.95 \\
    &= \$2565
\end{align*}

Out of which, the borrower already borrowed $2400$. So, an additional $\$165$ would be paid by the compound to the user. 

\textbf{Sale Price Determination:} The absorbed collateral is then made available for purchase at a price determined by the SFP:

\begin{align*}
    \text{Initial Sale Price} &= \text{Collateral Value} \times (1 - \text{LP} \times \text{SFP}) \\
    &= \$2700 \times (1 - 0.05 \times 0.60) \\
    &= \$2619
\end{align*}

\textbf{Post-Liquidation Sale:} If a liquidator buys the collateral at the sale price of $\$2619$, they can potentially sell it on the market for the current price of $\$2,700$, realizing a profit.
\\

\textit{Now, a more significant price drop post-liquidation can lead to insolvency if the collateral isn't sold quickly enough at the discounted price.}

\textbf{Post-Absorption Price Drop:} If the price of ETH drops further by $6\%$ in the next few blocks, before the sale, the new price would be:

\begin{align*}
    \text{Final Sale Price} &= \text{Initial Sale Price (at time of absorption)} \times (1 - \text{Percent Drop in Price}) \\
    &= \$2619 \times (1 - 0.06) \\
    &= \$2461.86
\end{align*}

\begin{align*}
    \text{Collateral Value (after Post-Absorption Price Drop)} &= \$2700 \times (1-0.06) \\
    &= \$2538
\end{align*}

\textbf{Protocol Loss:} At this point, the protocol has incurred protocol loss due to the shortfall in recovering the full amount owed by the borrower. The outstanding loss after liquidation is:

\begin{align*}
    \text{Protocol Loss} &= \text{Payment}-\text{Collateral Value} \times (1 - \text{LP} \times \text{SFP}) \\
    &= \$2565 - (2538* \times (1 - 0.05*0.6)) \\
    &= \$2565-2461.86 \\
    &= \$103.14
\end{align*}

\subsection{Risk Parameters Definitions}

\subsubsection{Borrow Collateral Factor}
The Borrow Collateral Factor quantifies the maximum permissible borrowing capacity for a user, expressed as a percentage of the total assets supplied within the protocol. Specifically, it dictates the upper bound of a user's borrowing capacity relative to the value of their supplied collateral. This parameter is critical in managing risk by ensuring that the value of borrowed assets does not exceed a safe threshold relative to the collateral provided.

\subsubsection{Liquidation Collateral Factor} 
The Liquidation Collateral Factor defines the ratio at which a user's loan becomes subject to liquidation. It is the ratio at which a borrower's position becomes undercollateralized and subject to liquidation. If the ratio of a borrower’s outstanding debt to the value of their collateral falls below this factor, the protocol can initiate liquidation to preserve its solvency. This metric is essential for maintaining the solvency of the protocol, as it delineates the boundary beyond which the collateral backing a loan is deemed insufficient.

\subsubsection{Liquidation Factor}
The Liquidation Factor defines the proportion of collateral that is returned to an account undergoing liquidation. Specifically, it determines the amount of collateral that is liquidated and paid out to the account in the event of liquidation. 

Conversely, the Liquidation Penalty, calculated as ($1- $Liquidation Factor), signifies the portion of the collateral seized by the protocol as a penalty for under collateralization. This penalty mechanism disincentivizes borrowers from allowing their positions to become liquidated, thereby contributing to the overall health of the protocol.

\subsubsection{Supply Cap}	
Supply cap refers to the maximum amount of an asset supplied to the protocol. Supply transactions will revert if the total supply is greater than this number as a result. For each collateral asset, there is a supply cap that limits the supply for each collateral asset for each chain. By implementing a supply cap, the protocol effectively mitigates the risks associated with collateral assets. This mechanism helps maintain the protocol’s balance and stability by preventing any single asset from dominating the reserves beyond a safe threshold.

\subsubsection{Storefront Price Factor}
The Storefront Price Factor denotes the fraction of the liquidation penalty allocated to buyers of collateral rather than the protocol itself. This factor plays a pivotal role in determining the discount rate applied to collateral available for sale during the account absorption process. 

The Storefront Price Factor plays a crucial role in incentivizing liquidators to engage in arbitrage and purchase the absorbed collateral at a discounted price. If the protocol sets the Storefront Price Factor too low, it may not be profitable for liquidators to participate, leading to a lack of liquidity in the market for absorbed collateral.

\subsubsection{Target Reserve}
The Target Reserve is the threshold reserve level of the base asset that, once attained, prohibits liquidators from purchasing collateral from the protocol. It represents the target amount of reserves of the base token. This parameter is instrumental in ensuring that the protocol maintains a sufficient reserve to manage its liabilities and operational requirements effectively.

\subsubsection{Liquidator’s Point}
In Compound V3, the protocol keeps track of liquidator points to tally the number of successful liquidations, or "absorbs", and the gas spent by the liquidator.  This points-based system is used to incentivize liquidators and manage risk. Liquidators earn points for successful liquidation calls and receive compensation for gas costs, which are then redeemable for rewards. This system helps ensure effective liquidation processes by aligning liquidator rewards with protocol efficiency and performance.

\subsubsection{Interest Rate Curve}
Compound V3 employs a dynamic interest rate model linked to asset utilization. Supply and borrow rates are determined independently by separate curves, each with a "kink" where rates sharply increase beyond a critical utilization threshold. This design helps balance supply and demand by adjusting interest rates dynamically based on real-time utilization.

Lenders earn interest, while borrowers incur charges, with rates reflecting current market conditions and accruing every second. This design enables the protocol to adjust rates dynamically, ensuring that they reflect current utilization and market conditions while maintaining balance and stability within the system.

\section{Simulated Stress Tests}

\subsection{Background of ABS}

\subsubsection{Agent-Based Simulations}

DeFi ecosystem is a balanced system that involves multiple agents having a close interaction amongst themselves \cite{jensen2021introduction}, \cite{werner2022sok}. The agent is an autonomous “entity” that can analyze its environment, which includes other existing agents, and interpret this data in any type of decision-making \cite{rutskiy2023dao}. Any agent can learn and improvise themselves over time by analyzing the existing data. To understand the workings of this ecosystem, it is important to understand each of these agents individually as well as the nature of their interactions. In DeFi, agents are mainly an emulation of users who interact in this DeFi ecosystem, which are mainly traders, arbitrageurs, liquidators, borrowers, liquidity providers, and more \cite{chen2023web3}.

\subsubsection{Monte Carlo Simulations}

Monte Carlo methods, or Monte Carlo experiments, are a broad class of computational algorithms that rely on repeated random sampling to obtain numerical results \cite{harrison2010introduction}. The underlying concept is to use randomness to solve problems that might be deterministic in principle. Monte Carlo is used in various real-life scenarios; for example, in financial markets, Monte Carlo simulations are used for long-term forecasting or understanding risk management in the respective market. They are also used in the medical field or to understand weather patterns over time.

\subsubsection{Why Agent-Based Monte Carlo Simulations}

An important class of applications for agent-based Monte Carlo simulations is financial applications, like estimating the distribution of the future value of a static portfolio \cite{jackel2002monte}. Here, we want to simulate protocol losses depending not only on price dynamics but also on the interaction of borrowers and liquidators. In each simulation, at every time step, we update what the borrowers and liquidators are doing. Because one simulation may not be able to give the complete picture, we will do multiple simulations to capture a broad range of possible outcomes. Indeed, so long as the simulations are statistically unbiased and independent, we may use such results to estimate the expected value of losses our protocols will sustain. Later in the report, we gave a detailed description of our methodology for the construction of these simulations, together with the statistical testing framework by which we could figure out the value at risk for several parameter configurations to come up with optimal parameter recommendations.

\subsubsection{Chainrisk Simulation Environment}

The Chainrisk Simulation Engine is a modular testing environment that is configured to align with the Compound’s existing state including lenders, borrowers, liquidators, contracts, and oracles. This engine has been used to run all the simulations and results in this report. The Chainrisk Simulation Engine is unique for its two-part high-fidelity simulation system.

\begin{itemize}
    \item RiskEVM - This is a Rust-based simulation engine optimized for Heavy Computations.
    \item On-Chain Simulation - This Engine is responsible for backtesting on forked networks and is optimized for factual precision.
\end{itemize}

The RiskEVM is a robust  Off-Chain, Rust based Simulation Engine that leverages Solidity as its development language. The term “Off-Chain” denotes the entirety of Compound contracts being deployed on a local network which allows for greater control and customization of various price models and risk parameters. This local network setup significantly optimizes testing, achieving approximately $40-45\%$ faster script execution compared to remote testnet tooling. The non-blockchain specifics are encapsulated away from the engine’s logic and can be fed in from an external source directly into the engine before execution starts. This provides the Data Analyst with the autonomy to design and structure the respective datasets as they see fit, within a general framework. Scripts are meticulously crafted  to facilitate the inclusion of external data. Due to Solidity’s unique numeric system, the external data being brought will be formatted correctly before being fed into the engine. This prevents any unforeseen errors due to conflicting numeric systems and is constant throughout the duration of the simulation. On completion of a particular simulation, the engine parses the resultant outcome to a standardized format and is stored for analysis. The RiskEVM is a Computation Optimized Engine i.e. it has been streamlined to handle a huge number of simulations while being highly fault tolerant.

\begin{figure}[]
     \centering
         \includegraphics[width=\textwidth, trim=0cm 0cm 0cm 0cm]{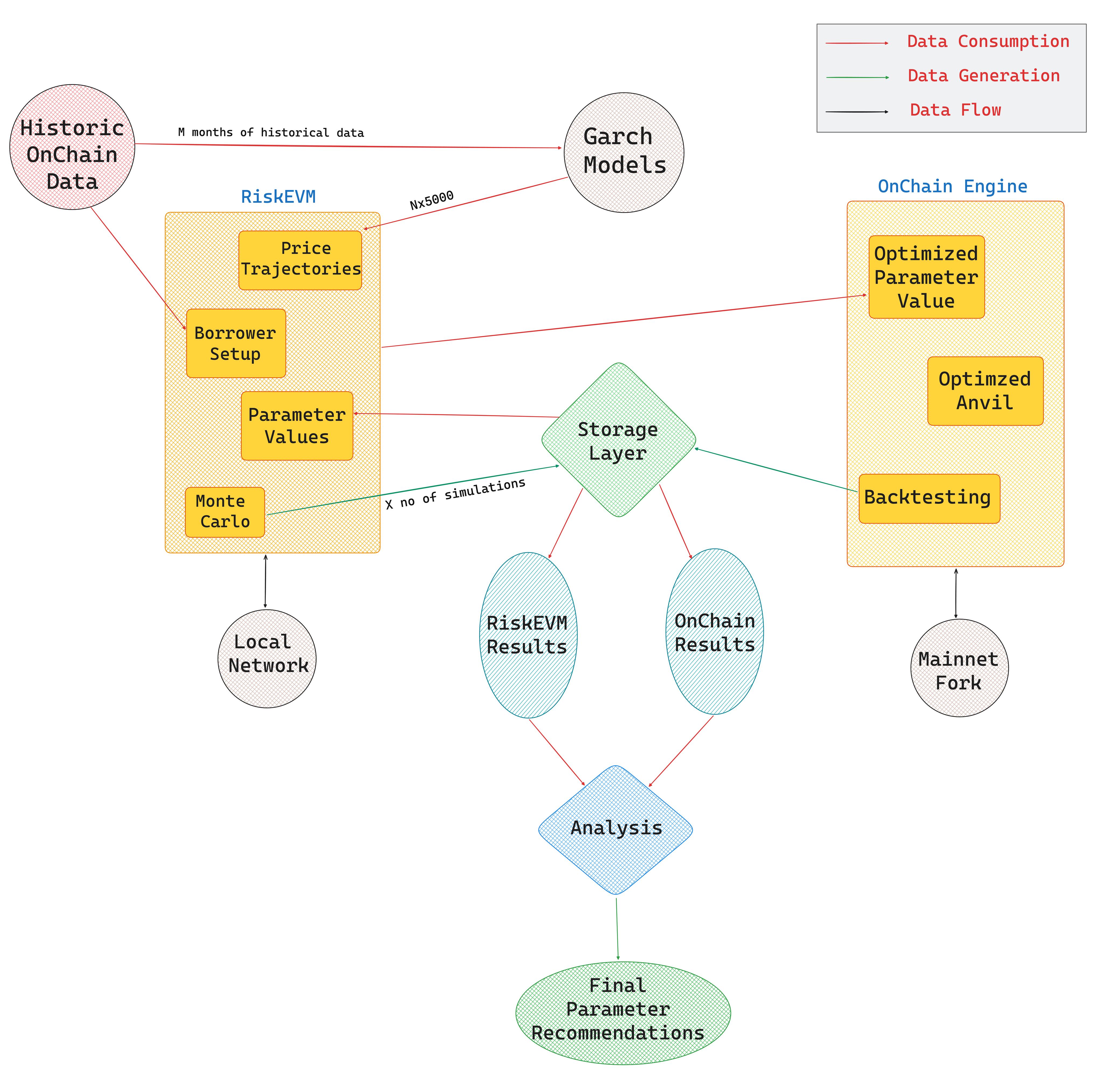}
         \caption{Chainrisk Simulation Engine Architecture}
\end{figure}

The On-Chain Engine is a custom simulation Engine built using Rust leveraging its highly performant and thread-safe concurrency. The term “On-Chain” denotes that this Engine operates on a Mainnet Fork mirroring the state of the real-world environment. The On-Chain engine is responsible for Backtesting the outcome of the RiskEVM, ensuring the simulations reflect actual network conditions and deliver precise, reliable results. The On-Chain Engine employs an optimized version of Anvil which has been configured to the requirements of the Chainrisk Simulation for better performance and high throughput. The transactions are executed on a real Mainnet fork which means the agent-based simulations are carried out exactly as it would in the real world.

\textbf{Why do we need 2 engines?}

The on-chain engine allows the agents to behave akin to the way they would in a production environment. Mainnet forks facilitate simulation on an exact snapshot of the blockchain at any given block height. This allows influencing factors to affect the execution of the transactions regardless of whether they are beneficial or detrimental to the user’s intent. Notwithstanding the obvious advantages of on-chain simulations, they come with their drawbacks as well. Running simulations on a DeFi protocol necessitates meticulous analysis with millions of scenarios being run with varying risk parameters. Moreover, mainnet forks rely on RPC calls which introduces latency and decreases overall runtime efficiency. Despite heavy optimizations, it is proven to be quite a resource-intensive and arduous task.

This is where the RiskEVM plays a crucial role, offloading the computation-heavy mission from the on-chain engine. The RiskEVM is optimized for running millions of simulations with their price trajectory and risk parameter values which can be changed on the fly. Moreover, the RiskEVM simulations are parallelized, enhancing throughput and performance.

\subsection{Risk Parameter Testing Methodology}

\subsubsection{Assumptions}

\begin{itemize}
    \item The Normality of Log Returns: The normal distribution characterizes the log returns of asset prices. This assumption simplifies statistical models and increases their tractability.
    \item Dependent Variable Is Normally Distributed: There is a normal distribution that guides slippage percentages, which makes it possible to use models assuming normality.
    \item Independence: The observations should be independent of each other. In other words, the price of an asset at any given time should not affect the price of an asset at any other given time in the future.
    \item Homogeneity: The asset should show a homogeneous nature about the price movement and overall reaction.
    \item Linearity: WBTC, ETH, and ARB have linear relationships between the sell variable and the slippage percentage transformed.
    \item Link Function: The link function chosen correctly relates the linear predictor to the mean of the distribution function, ensuring that there is an adequate capturing of the relationship among variables by the model.
    \item Static borrower: Borrower agents in the Chainrisk simulation environment are passive once initialized. Their initialization includes the supply and borrowing of collateral and base assets, respectively.
    \item Static liquidity: The protocol liquidity remains constant from external factors. This means outside of the delta produced by the borrower positions and liquidations; there is no change in the liquidity.
    \item Simulation Duration: We only look forward to a single day's worth of blocks.
    \item Price correlations: We assume that asset prices are correlated in real life and hence also in simulations.
    \item The Black Swan event is a statistical outlier since such an incident happened once to date since its inception. Thus, we decide not to test black swan events statistically. Moreover, the probability of such an event occurring is also pretty insignificant. Instead, we perform a separate “extreme VaR” methodology – which we will describe in greater detail in a subsequent paper.
\end{itemize}

\subsection{Agents}

\subsubsection{Price Trajectory}

\textbf{Introduction}

In any asset market, analyzing price trajectory is the backbone of mitigating losses in the market. Price trajectory is the expected path that a commodity (in this case, an asset) takes over a certain period. This path is sometimes heavily dependent on the historical price data, and finding a model that can extract the required amount of information from this historical data is a cumbersome job. Various statistical tools are available for this model building, but choosing the right tool requires many statistical tests. Here, we have tried to understand the nature of the available historical data with the help of various statistical tests and come up with a model henceforth. Compound protocol, from its very inception, has allowed the supply and borrowing of Ethereum (ETH) assets, providing a source of income for suppliers and helping borrowers at competitive rates. Thus, to assess the set of market risks that users of Compound protocol might face, the first step is to understand the price movements of the various assets on the Ethereum network. Section 5 illustrates the model used for developing the price path.
\vspace{5mm}

\textbf{Data Description}

The dataset includes blockwise data recorded every 200 blocks (approximately 50 seconds) for four assets on the Arbitrum Market on Compound - WBTC, ETH, GMX, and ARB. It captures transactions over 7 days, detailing the order size (sell value) and the resulting slippage percentage. The frequent data collection allows for a detailed analysis of market dynamics. After data cleaning, including the removal of outliers and consolidation of duplicates,the dataset was optimized for statistical analysis, ensuring the reliability and validity of subsequent findings. 

\vspace{5mm}
\textbf{Autoregressive (AR) Component}

In an Autoregressive (AR) model, future values of a time series are predicted based on its past values. Essentially, it assumes that the current value depends on a combination of its previous values. The AR model of order p (AR(p)) can be written as:
\[
X_t = \phi_1 X_{t-1} + \phi_2 X_{t-2} + \dots + \phi_p X_{t-p} + \epsilon_t
\]
Where:

\begin{itemize}
    \item \(X_t\) is the current value,
    \item \(X_{t-1}, X_{t-2}, \dots\) are past values,
    \item \(\phi_1, \phi_2, \dots\) are the weights for each past value,
    \item \(\epsilon_t\) is the error term.\end{itemize}
    
In simple terms, the AR model predicts the current value based on a weighted sum of past values.

\textbf{Moving Average (MA) Component} 

In an Moving Average(MA) model uses past errors (differences between actual and predicted values) to predict the current value. The MA model of order q (MA(q)) is written as:

\[
X_t = \theta_1 \epsilon_{t-1} + \theta_2 \epsilon_{t-2} + \dots + \theta_q \epsilon_{t-q} + \epsilon_t
\]

Here:
\begin{itemize}
    \item \(\epsilon_{t-1}, \epsilon_{t-2}, \dots\) are past errors,
    \item \(\theta_1, \theta_2, \dots\) are the weights for each error term.
\end{itemize}

The idea is that current values are influenced by previous errors.

\textbf{ARMA(p, q) model} 

This model combines both the AR and MA components, meaning it uses both past values and past errors to predict future values. The formula is:

\[
X_t = \sum_{i=1}^{p} \phi_i X_{t-i} + \sum_{j=1}^{q} \theta_j \epsilon_{t-j} + \epsilon_t
\]

ARMA models are used when a time series exhibits both autoregressive (values depend on past values) and moving average behavior (errors influence future values) \cite{guidolin2018autoregressive}. This model can capture both short-term fluctuations or correlations as well as long-term trends, making it a powerful tool for forecasting.

\textbf{GARCH}

GARCH, which stands for Generalized Autoregressive Conditional Heteroskedasticity, is a statistical model used in time series analysis to estimate and predict the volatility of financial returns \cite{ugurlu2014modeling}.

\textbf{Conditional Heteroskedasticity}
\begin{itemize}
    \item The term ``conditional heteroskedasticity'' means that the variance of the time series can change over time.
    \item GARCH models the variance at a given time as a function of past variances and past squared deviations from the mean (errors).
\end{itemize}

\textbf{Generalized}
\begin{itemize}
    \item The ``generalized'' aspect allows for a more flexible model compared to the original ARCH model.
    \item In ARCH, the current variance is modeled as a function of past squared residuals, while GARCH includes both past variances and past squared residuals.
\end{itemize} 
 
\textbf{GARCH Model Specification}

A GARCH(p, q) model is defined by the following equations:

\textbf{Mean Equation}
\begin{equation}
r_t = \mu + \epsilon_t
\end{equation}
where \( r_t \) is the return at time \( t \), \( \mu \) is the mean of the returns, and \( \epsilon_t \) is the error term.

\textbf{Variance Equation}
\begin{equation}
\sigma_t^2 = \alpha_0 + \sum_{i=1}^p \alpha_i \epsilon_{t-i}^2 + \sum_{j=1}^q \beta_j \sigma_{t-j}^2
\end{equation}
where \( \sigma_t^2 \) is the conditional variance at time \( t \), \( \alpha_0 \) is a constant, \( \alpha_i \) are the coefficients for the lagged squared residuals (ARCH terms), and \( \beta_j \) are the coefficients for the lagged variances (GARCH terms).
These are the mathematical formulas used to describe GARCH modeling, ensuring fairness in the functioning of the model to be implemented.

In light of the asset data, we came to the conclusion that we should use \textbf{GARCH}, the following are the reasons behind using this model.

\begin{itemize}
    \item \textbf{Captures Volatility Clustering:}
    \begin{itemize}
        \item Financial time series data as in here, such as asset prices and returns, often exhibit periods of high volatility followed by periods of low volatility giving an idea about the dynamic nature of market conditions and investor sentiment. This phenomenon is known as volatility clustering. GARCH models are effective in capturing this behavior, allowing for more accurate modeling and forecasting of price paths.
    \end{itemize}
    \item \textbf{Improving Forecast Accuracy:}
    \begin{itemize}
        \item By incorporating past information about both the asset returns and their variances, the GARCH model will provide more accurate forecasts of future volatility. This accuracy will be necessary for risk management.
    \end{itemize}
    
    \item \textbf{Option Pricing:}
    \begin{itemize}
        \item The most important aspect of using  GARCH models is they provide a way to estimate the conditional volatility, which can be used to price options more accurately. This is especially important for pricing options on assets with non-constant volatility which we have encountered.
    \end{itemize}
\end{itemize}

    
    

A correlated GARCH model, also known as a multivariate GARCH model, extends the univariate GARCH model to capture the dynamic correlations between multiple time series. This model is particularly useful in finance for analyzing the joint volatility of asset returns as it allows for the modeling of time-varying correlations that can reflect changing market conditions. By incorporating the correlation structure, the model provides a more comprehensive understanding of the co-movements and risk dynamics of the assets, making it a valuable model to be used in analyzing financial data.

\textbf{Computing price trajectories}

This section describes the development of the model for the price trajectories. In literature, two models are broadly used for price prediction and price trajectories: Stochastic Differential Equation (SDE) and Time series (also stochastic in nature). Existing Geometric Brownian Motion (GBM) models have been used for the Compound protocol \cite{gauntletreport2020compound}. In a simple GBM model, the price $(S)$ changes according to a stochastic differential equation: $dS_t = S_t(1 + \mu dt + \sigma dW_t)$, where $W_t$ is a Brownian motion, $\sigma$ is a volatility scaling parameter, and $\mu$ is a forcing parameter. However, in GBM models, the distribution and stationarity
of the underlying process are generally unrealistic when pricing data is modeled. Thus, we decided to use the time series model of GARCH. With the application of a battery of statistical tests on the historical price data, we decided on the order of GARCH and the corresponding ARMA model. The historical data has initially been subjected to testing of all the assumptions of the GARCH model and further statistical tests to develop the desired price prediction model.

\begin{figure}[h!]
     \centering
         \includegraphics[width=\textwidth, trim=0cm 0cm 0cm 0cm]{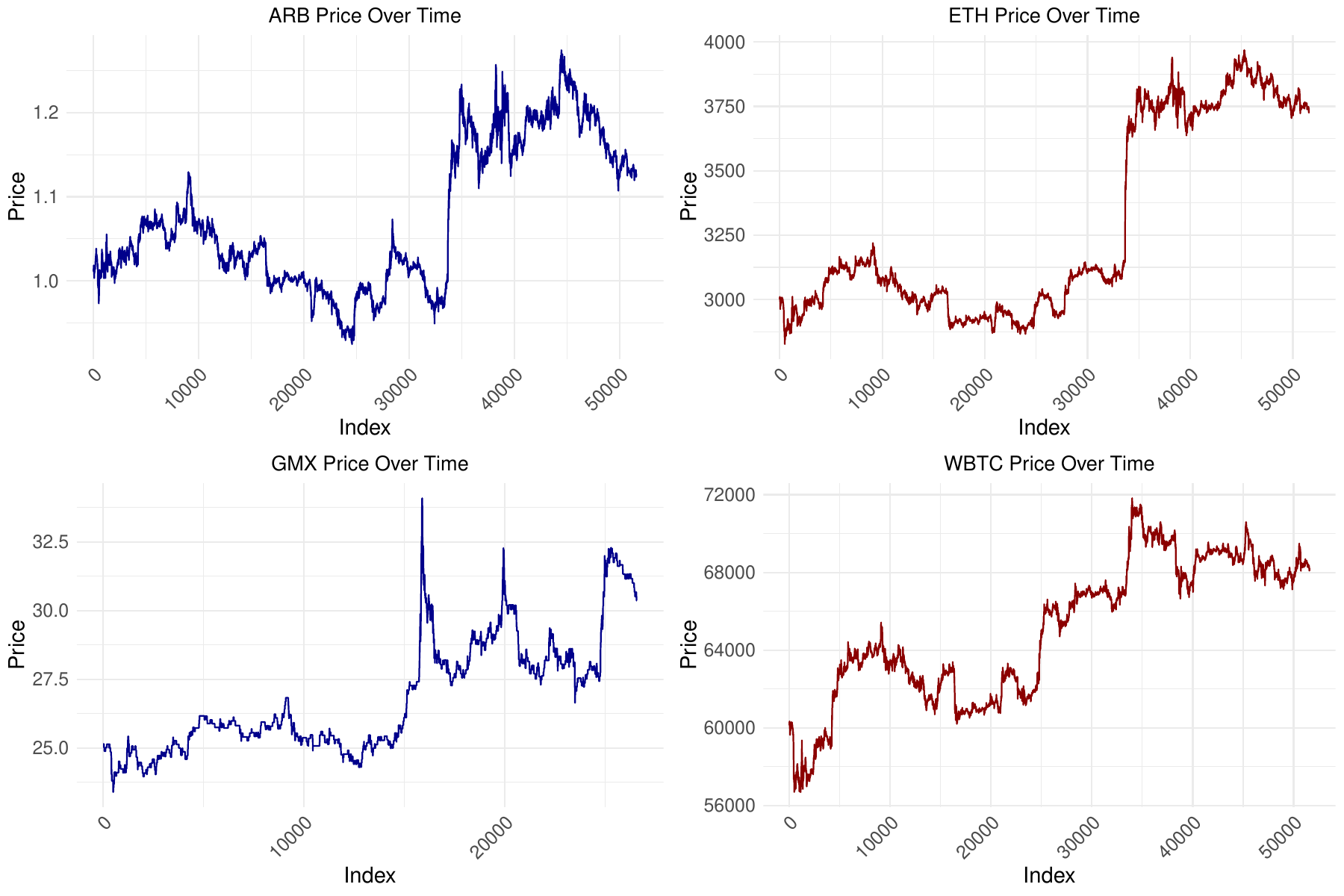}
         \caption{ Asset-wise price (per minute) for one month}
        \label{fig:historic price}
\end{figure}

We see two scenarios, namely single-asset and multi-asset price trajectories. In a single-asset price trajectory, we have decided to observe the historical price movement of Ethereum for the last month, with the price taken every minute. The price movement that we observed in the historical data is observed in the second subfigure of the first row of Figure \ref{fig:historic price}. This historical price data is used to obtain single-asset price modeling.

Further, for the multi-asset scenario, we have taken the idea of a multivariate normal distribution to obtain the multi-asset model. In the multi-asset scenario, we have to initially understand the dependence of one of the assets on the remaining assets. The main work is to assume the model that will be taken as a base model (one for which the covariance of remaining assets will be observed). Until now, no concrete methodology has been used to decide the base asset. A subjective approach was taken to decide the base model in our scenario. Here, we have assumed the base model to be Ethereum. Depending on the chosen base model, the covariance of the remaining assets has been observed, and we use the idea of multivariate normal to obtain the dependent price trajectories of the remaining assets along with the base asset.

\textbf{Simulation}

The historical data used for the simulation is sourced from different assets, but it is uniformly sampled at intervals of 200 blocks over the past month. This consistent sampling interval ensures that the data is comparable across different assets and allows for a standardized analysis period. Initially, we employed a single-asset model using Ethereum's historical data. This model demonstrated a lower Mean Squared Error (MSE) compared to other models when tested across various test-split ratios. Additionally, we evaluated other efficiency statistics such as the Akaike Information Criterion (AIC) and Bayesian Information Criterion (BIC) to assist in selecting the most effective model.

\begin{figure}[h!]
     \centering
         \includegraphics[width=\textwidth, trim=0cm 1cm 0cm 0cm]{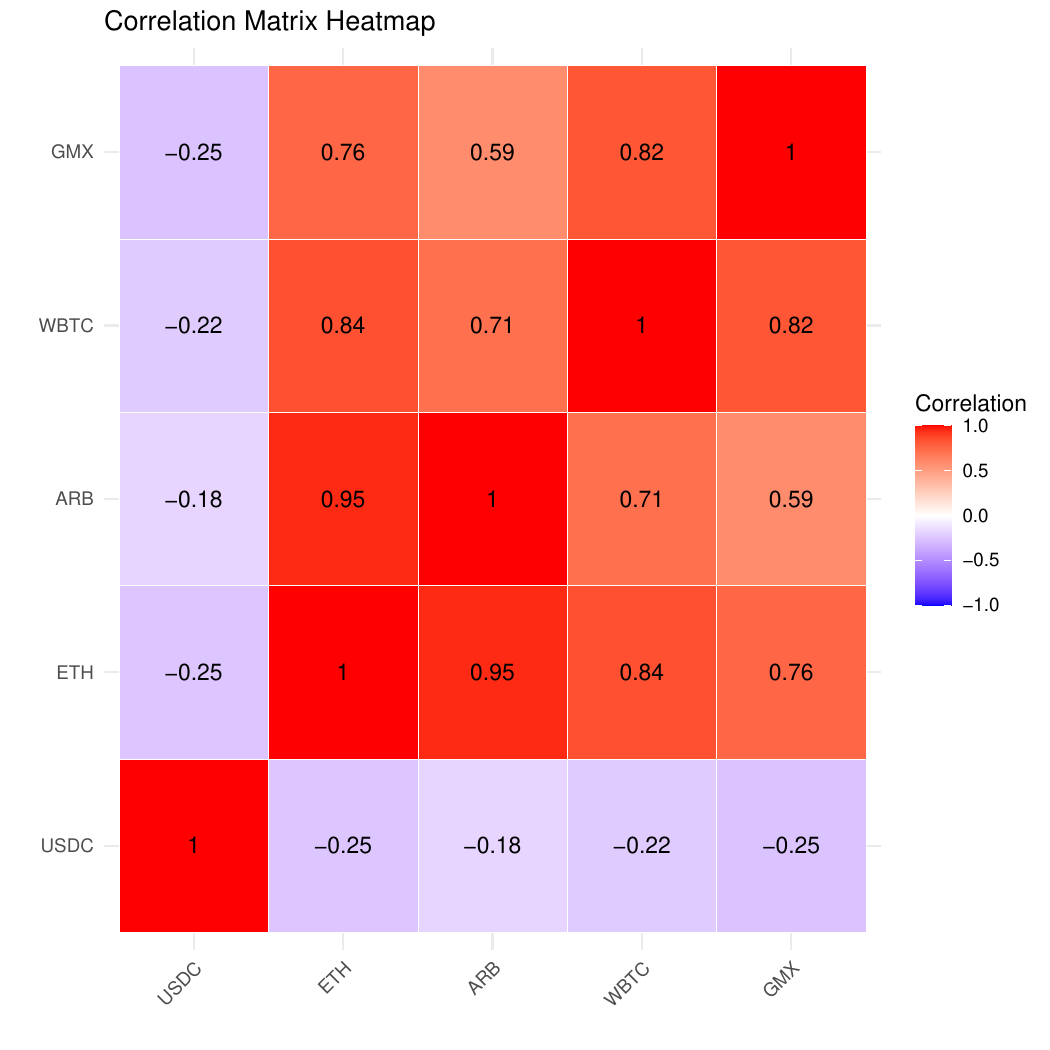}
         \caption{Correlation Heat-map for prices (per minute) of 5 assets taken over one month}
        \label{fig:multiasset_corr_heatmap}
\end{figure}

In the subsequent phase, we expanded our approach to a multi-asset scenario by accounting for the interdependencies among the price movements of different assets. The historical data revealed a significant correlation structure between the assets under consideration, indicating that their price movements are not entirely independent. Leveraging this correlation structure, we modeled the multi-asset price trajectory.

The base asset, Ethereum, was chosen subjectively based on its market significance and data availability. The correlation 
structures of the other assets were then analyzed in relation to Ethereum. This step involved calculating the correlation coefficients and creating a correlation matrix to quantify the relationships between the asset prices.

Using this information, we developed a multi-asset price trajectory model by incorporating the correlation matrix and applying the corresponding Generalized Autoregressive Conditional Heteroskedasticity (GARCH) model with the selected parameters. The GARCH model was chosen for its ability to model volatility clustering and time-varying volatility, which are common characteristics in financial time series data. By including the correlation structure, the model accounts for the dependencies among the assets, allowing for a more accurate representation of their joint dynamics.

\textbf{Overview of Simulated Price Paths}

\begin{figure}[h!]
     \centering
         \includegraphics[width=\textwidth, trim=0cm 4cm 0cm 1cm]{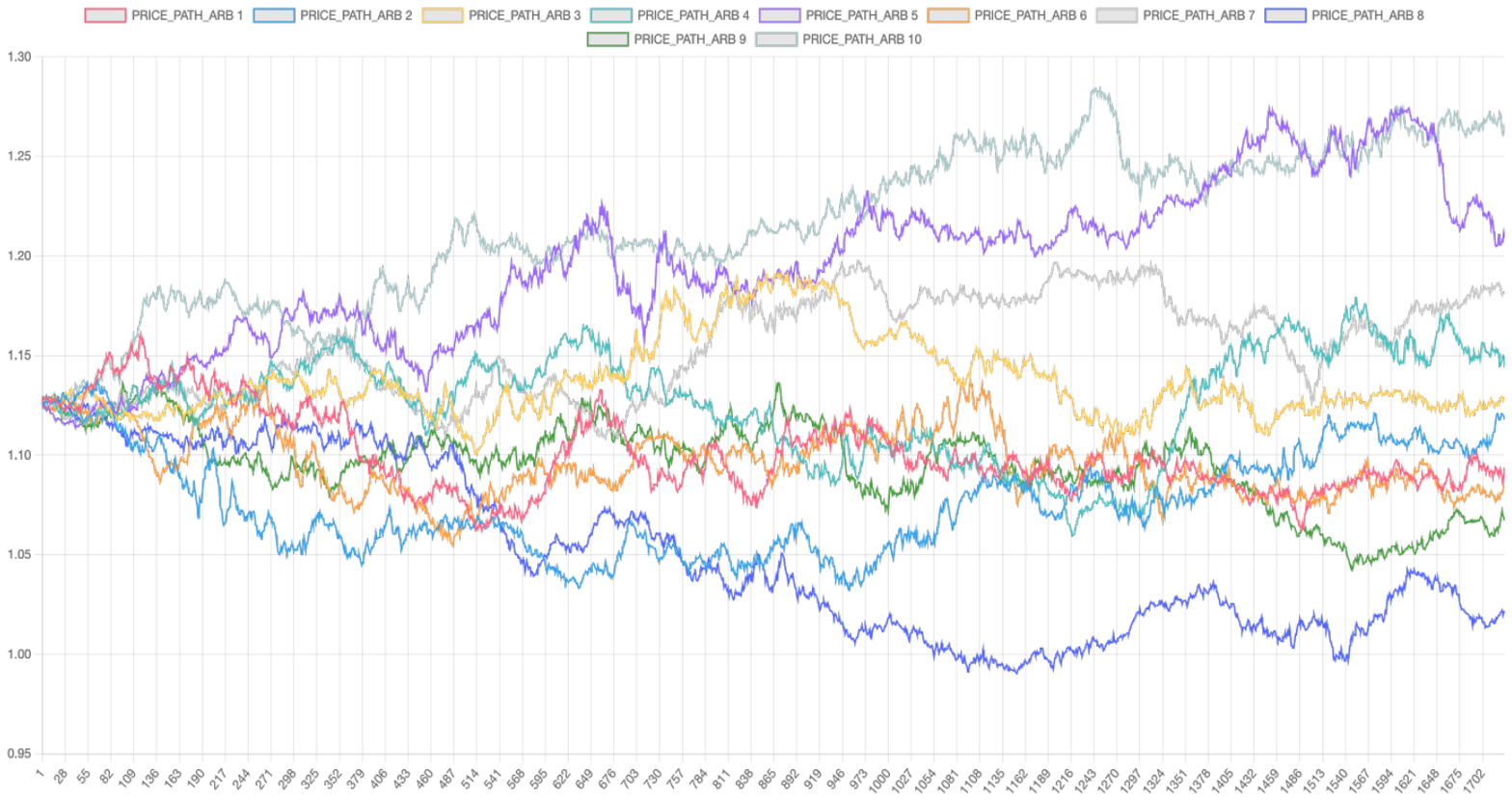}
         \caption{Arbitrum Price Paths for 10 paths}
        \label{fig:Arbitrium Prices}
\end{figure}

\textbf{Arbitrum}
Figure \ref{fig:Arbitrium Prices} exhibits the arbitrium price paths showing the variation of 10 different paths and also capturing the volatility. 
Arbitrum, being a newer and smaller asset compared to the others, as expected exhibits overall higher volatility. This is common in less mature markets where price swings can be more pronounced due to lower liquidity and higher speculative interest. The maximum volatility is 15.666%

\begin{figure}[h!]
     \centering
         \includegraphics[width=\textwidth, trim=0cm 4cm 0cm 1cm]{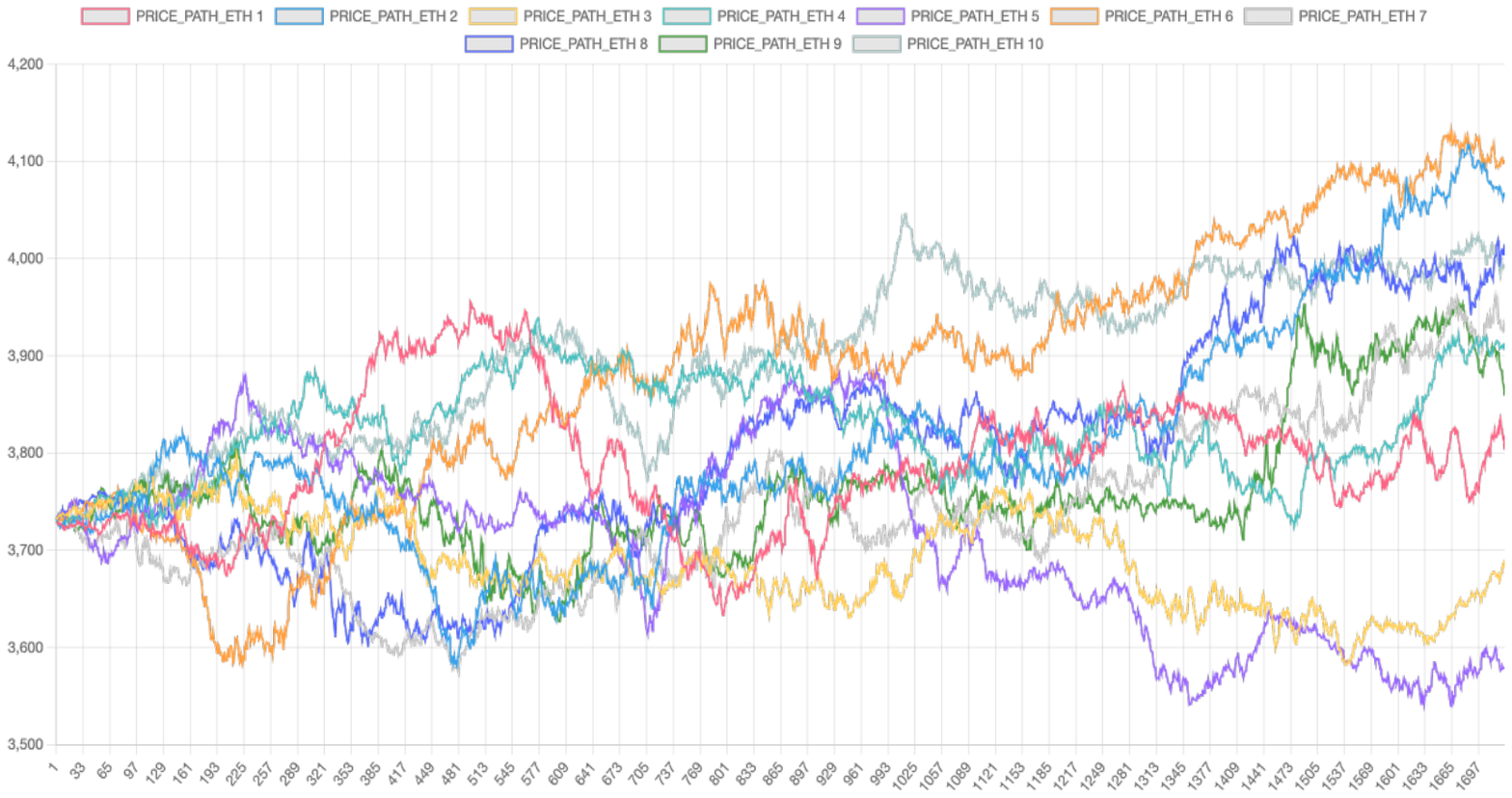}
         \caption{Ethereum Price Paths for 10 paths}
        \label{fig:Ethereum Prices}
\end{figure}

\textbf{Ethereum}
Figure \ref{fig:Ethereum Prices} exhibits the Ethereum price paths showing the variation of 10 different paths and also capturing the volatility.
Ethereum, being a major cryptocurrency, is showing substantial but somewhat stable volatility. Its price movements influenced by market trends, network upgrades, and broader adoption of decentralized applications and smart contracts. The maximum volatility that is been exhibited is of price path 3 which is 9.13

\begin{figure}[h!]
     \centering
         \includegraphics[width=\textwidth, trim=0cm 4cm 0cm 1cm]{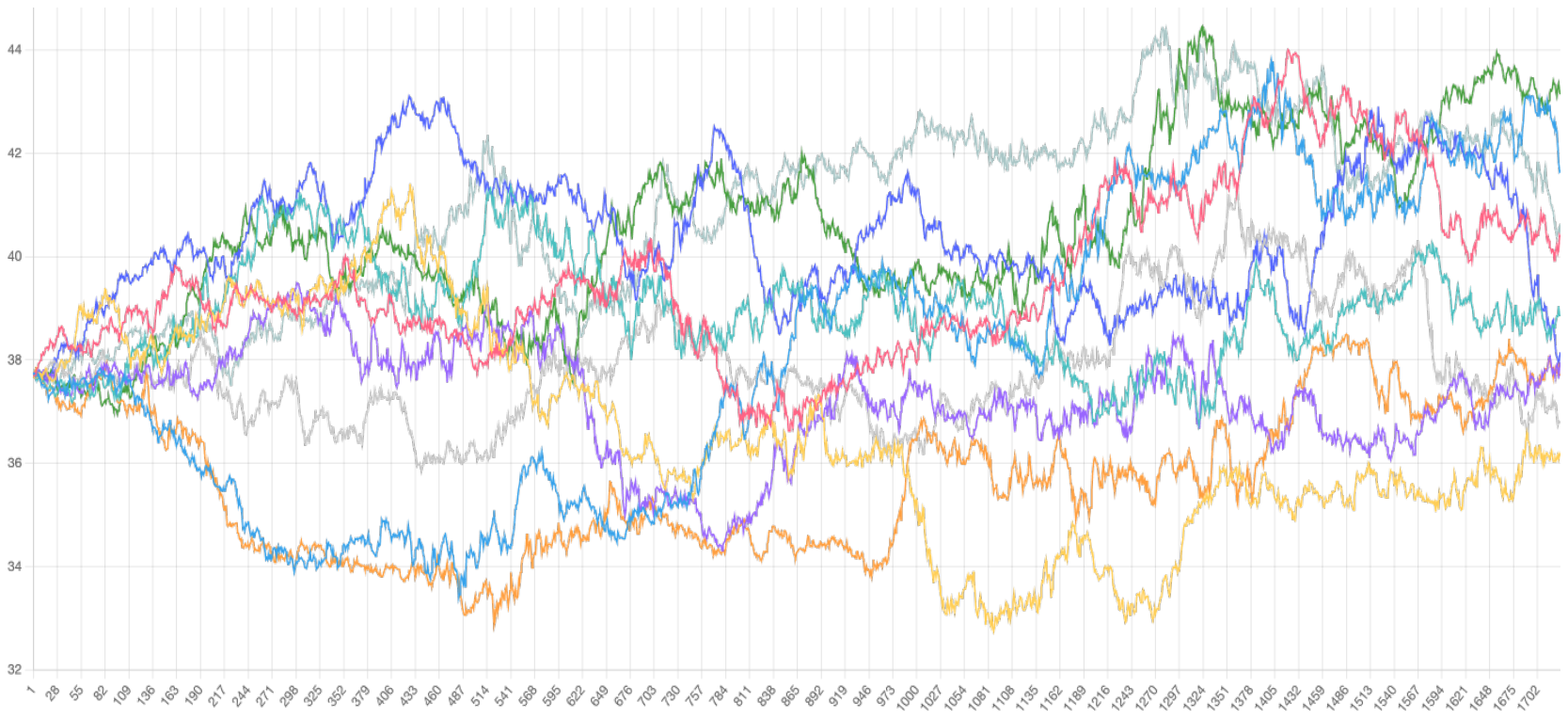}
         \caption{GMX Price Paths for 10 paths}
        \label{fig:GMX Prices}
\end{figure}

\textbf{GMX}
Figure \ref{fig:GMX Prices} exhibits the  GMX price paths showing the variation of 10 different paths and also capturing the volatility.
This decentralized exchange token shows moderate to high volatility. As it's tied to the performance and adoption of the GMX platform, its price can be influenced by platform usage metrics, broader market conditions, and speculative trading.
Although the overall volatility is moderate there is an extreme condition that is reflected by the eighth price path giving the maximum volatility to be 41.33 percent.

\begin{figure}[h!]
     \centering
         \includegraphics[width=\textwidth, trim=0cm 4cm 0cm 1cm]{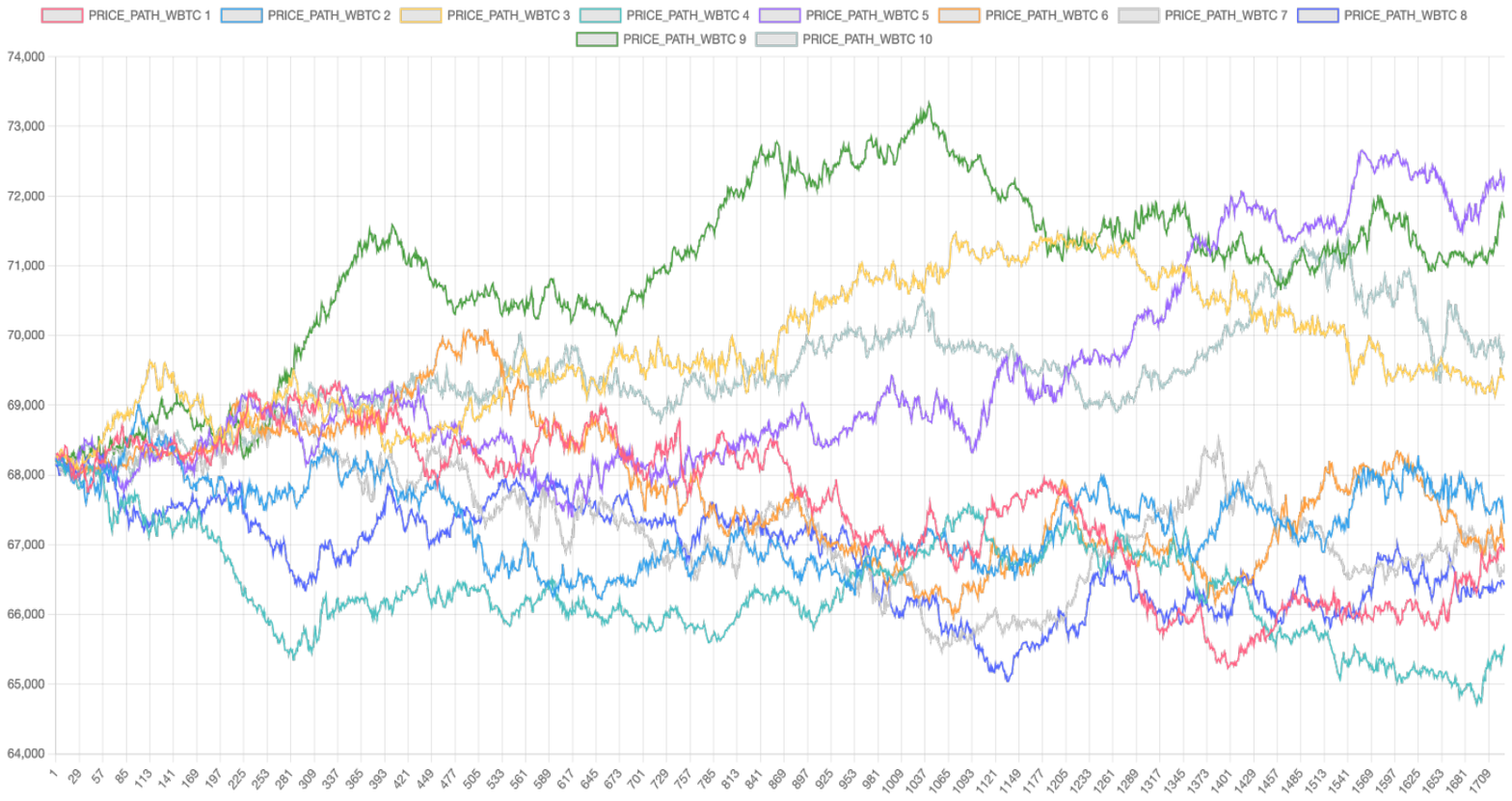}
         \caption{WBTC Price Paths for 10 paths}
        \label{fig:WBTC Prices}
\end{figure}

\textbf{WBTC}
Figure \ref{fig:WBTC Prices} exhibits the  WBTC price paths, showing the variation of 10 different paths and also capturing the volatility.
WBTC, being pegged to Bitcoin, should exhibit volatility similar to Bitcoin. Here, WBTC exhibits an overall low volatility.
The maximum volatility is exhibited in the second price path of 4.4 percent.

\textbf{Percentage change in first price path}

\begin{figure}[h!]
     \centering
         \includegraphics[width=\textwidth, trim=0cm 18.5cm 0cm 0.25cm]{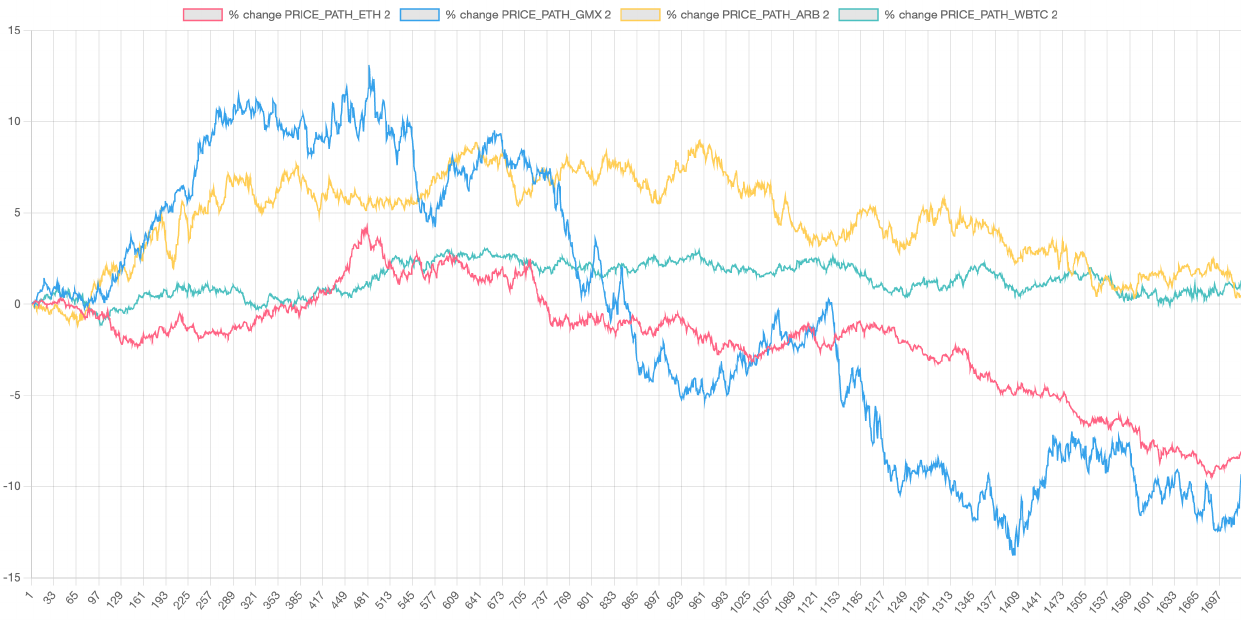}
         \caption{Percentage change in one out of all simulated price paths}
        \label{fig:PricePath2percentchange}
\end{figure}

Figure \ref{fig:PricePath2percentchange} exhibits the percentage change in all assets.
GMX and Ethereum show the maximum percentage change although the GMX prices have an increasing trend and Ethereum has a decreasing trend in the percentage change. It can be seen that while Ethereum has observed a price drop of nearly $-8\%$ while on the other hand, GMX has seen an increase of $8\%$.
Arbitrum shows a moderate decreasing trend with a final decrease of $-5\%$, with WBTC showing the least overall movement for that specific day of simulation. This is one of the scenarios of all the simulated price paths. 

Financial simulations allow us to evaluate the potential price movements of various assets under different market conditions, offering insights into risk and reward. In each simulation set, we track the maximum drop (in $\%$) (worst case) and maximum increase (in $\%$) (best case) for assets like ETH, GMX, ARB, and WBTC. The maximum drop reflects the worst-case scenario, showing how much the price can fall from its initial value, while the maximum increase highlights the best-case scenario, indicating the greatest potential price surge. These percentages give us an idea of how volatile each asset can be, helping investors prepare for both downturns and opportunities for profit.

In the first simulation set, ETH saw a drop of $19\%$ and an increase of $20\%$, GMX experienced a drop of $28\%$ and an increase of $29\%$, ARB dropped by $30\%$ and rose by $31\%$, and WBTC showed a decline of $18\%$ and a rise of $20\%$. All other simulated price paths have a price drop or increase within this range, thus spanning a significantly large interval with length of  nearly $40\%$. Similarly, for other sets, the maximum price drop and increase are tabulated in Table \ref{tab:price_change}. These simulated scenarios allow investors to better understand the range of possible outcomes for each asset and make informed decisions. Higher volatility means greater risk but also the possibility of higher rewards. By analyzing different simulation sets, investors can compare how each asset might perform under varying conditions, helping to craft strategies for risk management, profit-taking, and long-term investment planning.

\begin{table}[h!]
\centering
\caption{Price Change (in $\%$) for the three sets of simulations}
\resizebox{\textwidth}{!}{
\arrayrulecolor{black}
\begin{tabular}{|c|c|c|c|c|}
\hline
\multirow{2}{*}{\textbf{Simulation Set}} & \textbf{ETH price} & \textbf{GMX price} & \textbf{ARB price} & \textbf{WBTC price} \\
& \textbf{(drop\%, increase \%)} & \textbf{(drop\%, increase \%)} & \textbf{(drop\%, increase \%)} & \textbf{(drop\%, increase \%)}\\
\hline
\rowcolor{yellow!20}
1 & (-19, 20) & (-28, 29) & (-30, 31) & (-18, 20) \\
\hline
\rowcolor{green!20}
2 & (-38, 29) & (-36, 32) & (-29, 29) & (-22, 13) \\
\hline
\rowcolor{yellow!20}
3 & (-20, 27) & (-33, 31) & (-26, 32) & (-18, 16) \\
\hline
\end{tabular}
}
\label{tab:price_change}
\end{table}

\subsubsection{Borrower Agents}

The Borrower Agents are initialized based on on-chain historical data on Arbitrum Mainnet. We take into consideration only significant borrowers with a minimum borrowed amount. In our case, the minimum borrowed amount is $\$1000$. Additionally, there is an upper bound on the Health Factor of the borrower's address. The Health Factor must be equal to or lower than 2. This criterion ensures that only positions at risk of becoming underwater due to adverse price changes are considered. Borrower Agents in the Chainrisk Simulation Environment are passive, once initialized. They are initialized at the start of the simulation, following which their state(balance) remains unchanged unless there is a liquidation event on their position. 

\subsubsection{Slippage}
Building on the understanding of price trajectories, we now delve into a comprehensive analysis of slippage data to explore the relationship between slippage and trading activities. This section presents the methodologies, models, and results from our detailed analysis.

\begin{itemize}
    \item \textbf{Data Loading}: The datasets were loaded from JSON files and include information for four assets: WBTC, ETH, GMX, and ARB. The dependent variable in this analysis is the slippage percentage, while the independent variable is the order size, represented by the transaction's sell value. The initial analysis involved plotting graphs to visualize the relationship between sell values and slippage percentage. This helped to understand the data distribution and identify any anomalies. The dataset captures transactions over a duration of 7 days, with data points recorded each time a transaction was triggered.
    
    \item \textbf{Data Cleaning}: Several data cleaning steps were undertaken to ensure the integrity and quality of the dataset. First, influential points identified in the initial graphs were removed to provide a clearer understanding of the relationship between sell values and slippage percentage. These influential points were such that they would either not affect the analysis on an overall scale or they would not lead to liquidation. There were scenarios where the slippage percentage has been negative, implying there has been no slippage loss, which will not lead to any issue in liquidation, so those points have been removed, as they might be influential points for the analysis. Similarly, very few whale accounts were removed as they were also influential points for the analysis. Next, duplicate entries with identical sell values and slippage percentages were consolidated into a single entry to avoid redundancy. These steps helped refine the dataset, making it more accurate and reliable for further analysis.
    
    \item \textbf{Density function of Slippage}: To further understand the distribution of slippage percentage, its density function was plotted. This helps to visualize the probability distribution of slippage percentage, providing insights into its central tendency, variability, and potential skewness. It can be seen in the graph that the distribution was more or less normal ( which was the same for other assets with different means). So, we can use the models with the assumption of a normal dependent variable.

    \item \textbf{Final Model Selection}: For WBTC, ETH, and ARB, we used log-transformed values, while for GMX, we opted for a simple GLM model. Initially, graphs were plotted using the original 'sell' variable, but the results were unsatisfactory. A linear regression model was fitted, yielding poor results except GMX, which was made based on the p-value obtained using a simple linear regression model. To improve the model's performance, a logarithmic transformation was applied to the 'sell' variable for all assets except GMX, making the relationship between the independent and dependent variables more linear.
    
    \begin{figure}[]
     \centering
         \includegraphics[width=\textwidth, trim=0cm 0cm 0cm 1.2cm]{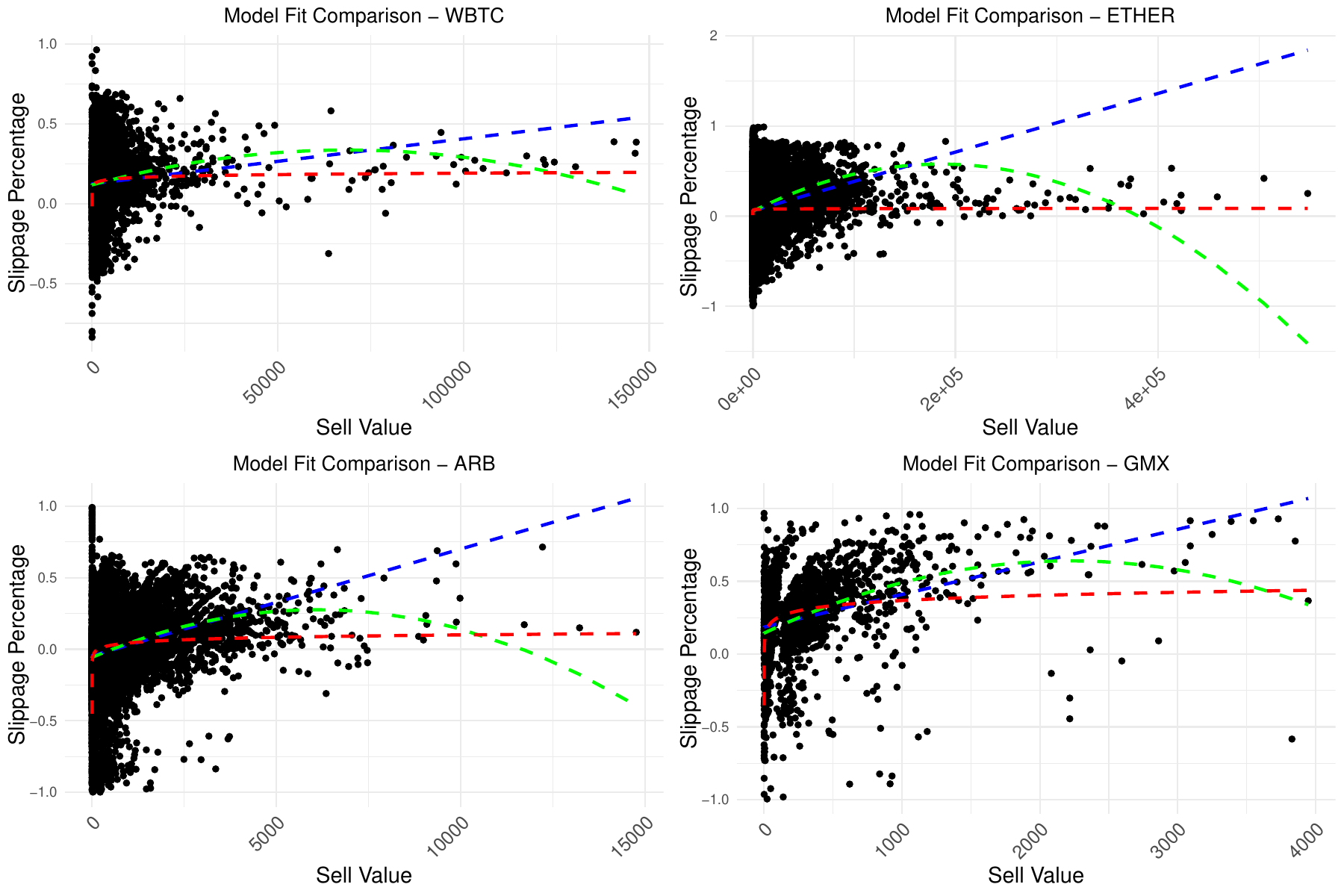}
         \caption{Scatter plot with three different regression lines}
        \label{fig:scatterplotwithreglines}
    \end{figure}

    In the Figure \ref{fig:scatterplotwithreglines}, the blue line depicts the linear model, the red line depicts the log-transformed model, and the green line depicts the quadratic model.

    \item \textbf{Model Fitting and Evaluation}: To evaluate the model's performance further, cross-validation was performed to determine the Root Mean Squared Error (RMSE) and Mean Absolute Error (MAE). The results of model fitting are as follows:

\end{itemize}
\begin{table}[h!]
\centering
\caption{Results}
\resizebox{\textwidth}{!}{
\begin{tabular}{|c|c|c|c|c|}
    \hline
    \textbf{Asset} & \textbf{Slope Equation} & \textbf{AIC} & \textbf{RMSE} & \textbf{MAE} \\
    \hline
    WBTC & y = 0.0421 + 0.0129 * logSell & -20249.26 & 0.15 & 0.11 \\
    \hline
    ETH & y = 0.057 + 0.0023 * logSell & -178868.85 & 0.17 & 0.12 \\
    \hline
    ARB & y = -0.124 + 0.0244 * logSell & -13234.21 & 0.19 & 0.12 \\
    \hline
    GMX & y = 0.186 + 2e-04 * sell & 432.84 & 0.27 & 0.20 \\
    \hline
\end{tabular}
}
\end{table}

\begin{figure}[h!]
     \centering
         \includegraphics[width=\textwidth, trim=0cm 0.5cm 0cm 0.325cm]{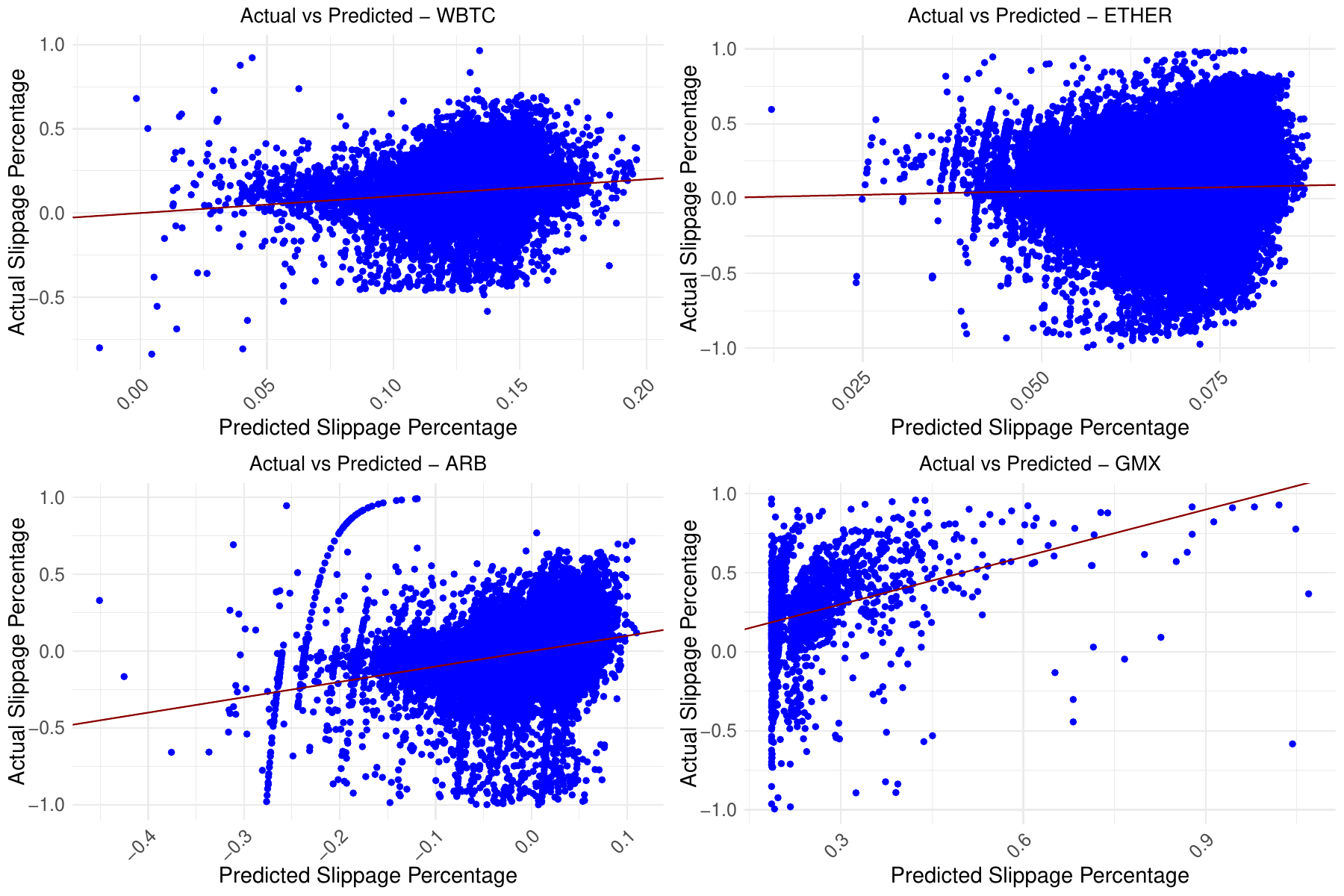}
         \caption{Scatter plot of Actual versus predicted slippage values}
\end{figure}

\subsubsection{Liquidator Agents}
Liquidator agents play a crucial role in the liquidation process within decentralized finance protocols. They operate based on a profit function that is influenced by several key parameters, which dictate their behavior during liquidations.

Key Parameters Influencing Liquidator Behavior - 

\begin{itemize}
    \item Slippage of Collateral Asset
    \item Trading Fee on DEXes
    \item Liquidation Penalty
    \item StoreFront Price Factor
\end{itemize}
Note: Gas fees have been excluded from this analysis. Historical data for gas prices on the Arbitrum chain indicated minimal fluctuations that would significantly impact liquidator behavior. For trading fees, a standardized value of $0.3\%$ was adopted based on Uniswap's typical fee structure. Slippage was modeled using historical data to accurately represent real-world conditions.

\begin{figure}[h!]
     \centering
         \includegraphics[width=\textwidth, trim=0cm 0cm 0cm 0cm]{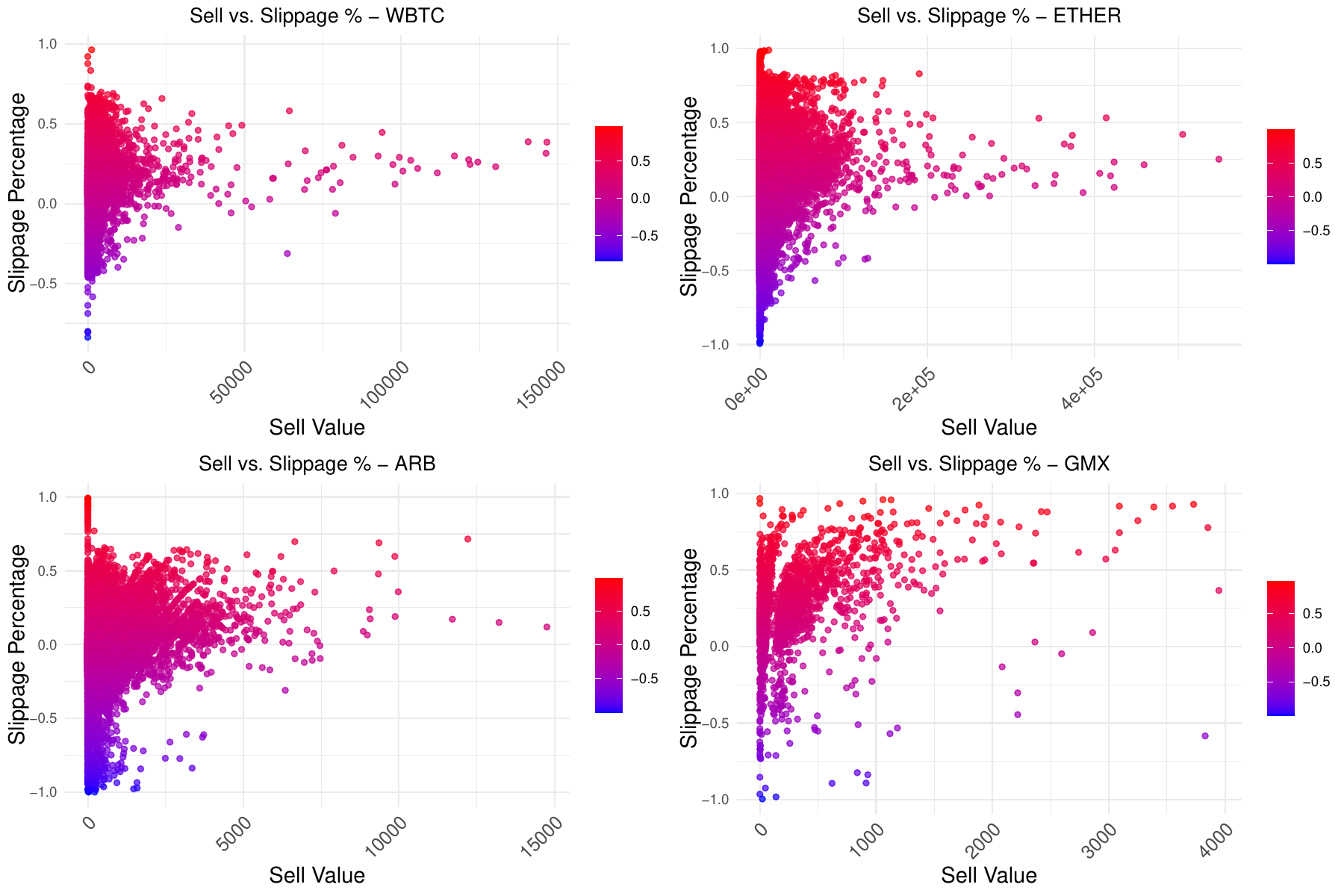}
         \caption{Scatter plot of Sell values versus slippage (in $\%$)}
\end{figure}

\textbf{Condition for Liquidator Bot to Buy Collateral}

To determine whether the liquidator bot should proceed with purchasing collateral, the following condition must be satisfied:

\begin{equation*}
    Max\ Liquidator\ Slippage (\%) \leq  LP * SFP
\end{equation*}

\textbf{Profit Function for Liquidator}
The profit function for the liquidator agent can be expressed as follows:

\begin{equation*}
    P = S_C-P_C-(f \times S_C)-S
\end{equation*}
where
$P$ represents profit, $S_C$ represents the sale price of the Collateral, $P_C$ is purchase price of the collateral, $f$ is the Trading fee rate, $S$ represents the slippage costs.

\begin{figure}[h!]
     \centering
         \includegraphics[width=\textwidth, trim=0cm 0cm 0cm 0cm]{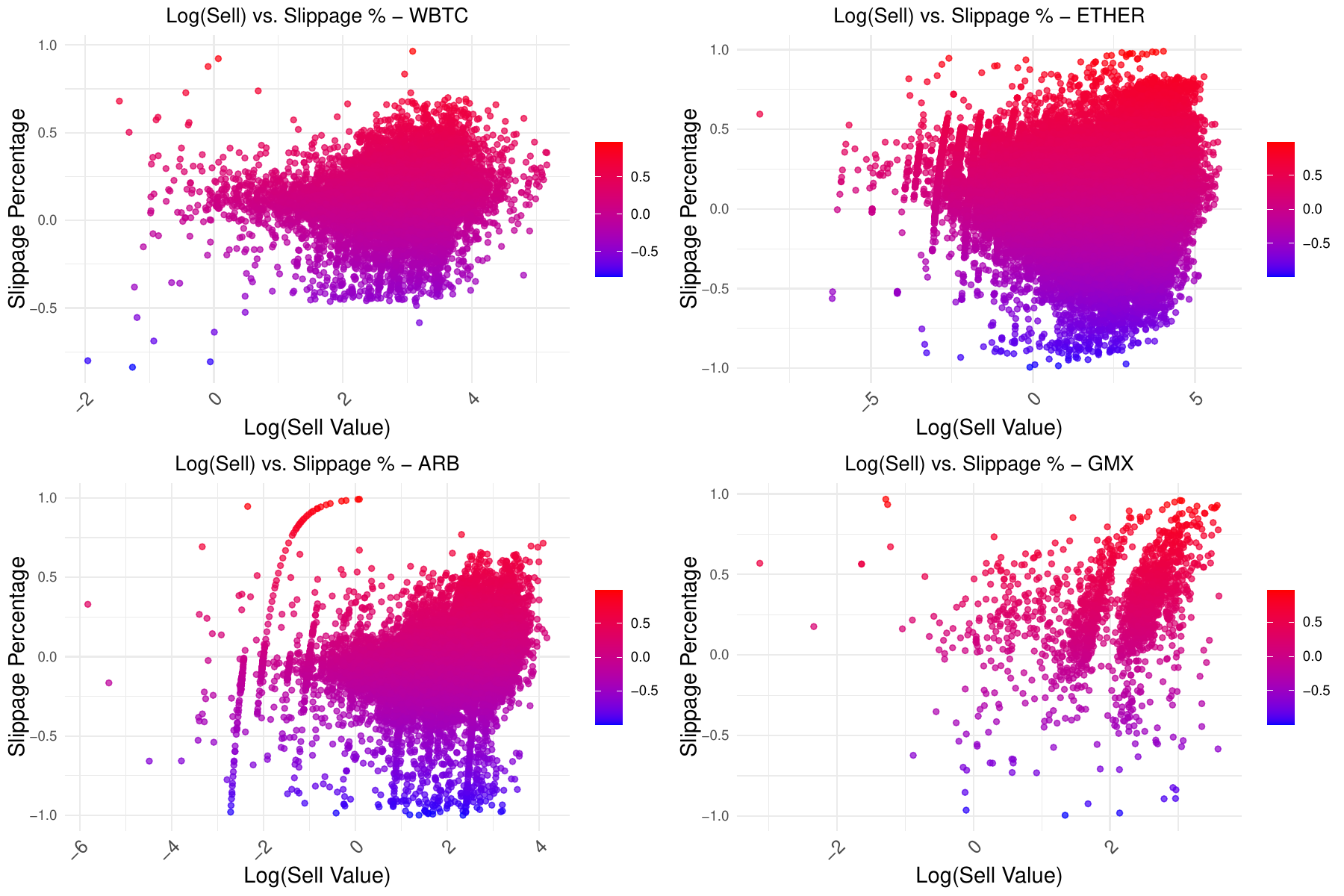}
         \caption{Scatter plot of Slippage (in $\%$) versus the logarithm of sell values}
\end{figure}

\textbf{Liquidator Bot Logic}

The logic of the liquidator bot operates as follows:
\begin{itemize}
    \item Check Conditions: The bot first verifies if the slippage and trading fee are less than the product of LP and SFP expressed in percentage terms.
    \item Fetch Collateral Value: If the conditions are satisfied, the bot fetches the value of the available collateral for purchase in the base asset.
    \item Borrow Base Token: The bot borrows the necessary base token from a designated Uniswap pool using the flash swap functionality.
    \item Buy Discounted Collateral: The bot then purchases the discounted collateral from the protocol.
    \item Exchange Collateral Assets: After acquiring the collateral, the bot exchanges the collateral assets into the base token using other Uniswap pools.
    \item Repay Flash Loan: Finally, the bot pays back the flash loan and sends the profit in the base token to the caller of the liquidator contract.
\end{itemize}

\begin{figure}[h!]
     \centering
         \includegraphics[width=\textwidth, trim=0cm 0cm 0cm 0cm]{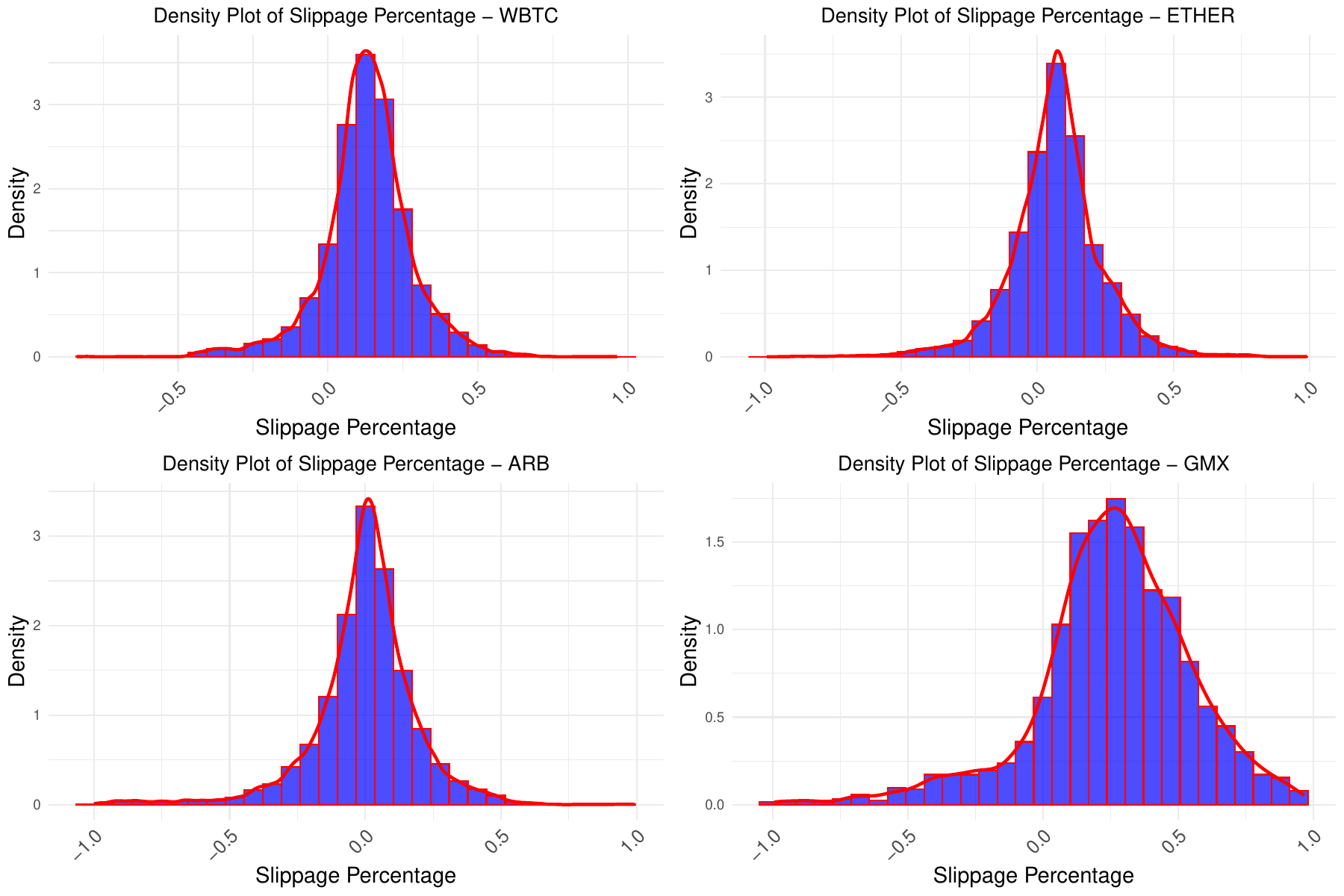}
         \caption{Density plot of Slippage (in $\%$) asset-wise}
\end{figure}

\subsubsection{Liquidity Supply}

In our simulation, a supplier contributes USDC to the protocol as the base asset, which begins accruing interest. The supplier receives a corresponding quantity of cUSDCv3, equal to the amount of USDC supplied. This supplier role is crucial as it provides liquidity to the system, enabling borrowers to withdraw USDC from the protocol. 
We took on-chain snapshots of all the liquidity suppliers and replicated this exact on-chain data within our simulation. This ensures that our model accurately reflects the current state of liquidity.

\subsection{Estimating Value at Risk}
To estimate the Value at Risk (VaR) using a simulation-based approach, the following steps are taken:
\begin{itemize}
    \item Initial Simulation: $5000$ simulations were run using an initial set of estimated parameters $s$. Once these simulations are complete, calculate the $95^{th}$ percentile of the losses generated from these $5000$ iterations. This percentile represents the potential loss level that will only be exceeded $5\%$ of the time.
    \item Convergence Check: To ensure the reliability of the $95^{th}$ percentile estimate, run an additional $5000$ simulations using the same parameters, $s$. Combine these new simulations with the initial set to create a total of $10,000$ simulations. Calculate the $95^{th}$ percentile loss again based on this larger dataset. Compare this new $95^{th}$ percentile with the one obtained from the initial 5000 simulations. If the absolute difference between these two percentiles is within a pre-specified tolerance level $s$, the process can move forward. If not, this step is repeated until the difference is within the acceptable bound.
    \item Final Calculation: Once the convergence condition is met, conduct a final round of $5000$ simulations under the same parameters, bringing the total to $15,000$ simulations. If the difference in $95^{th}$ percentiles from previous steps is still within $\epsilon$, the final VaR is estimated as the $95^{th}$ percentile loss from the entire set of $15,000$ simulations. This ensures that the VaR calculation is both stable and accurate.
\end{itemize}

\subsection{Calculating Protocol Losses}
    To determine the total protocol losses for a given period, we aggregate the losses incurred from each individual position. It is essential to note that most positions maintain a health factor above $1$ and, therefore, do not contribute to protocol losses.

    Liquidation occurs in two distinct steps:

    \begin{itemize}
        \item \textbf{absorb()}: In this step, the protocol absorbs the underwater position. The protocol settles the user’s debt using its base asset reserve. Once the debt is cleared, the collateral is transferred to the protocol's collateral reserve, making it available for sale.
        \item \textbf{buyCollateral()}: During this step, the liquidator has the opportunity to purchase the absorbed collateral from the protocol at a discounted rate.
    \end{itemize}

A potential loss scenario arises when the collateral's market value depreciates after absorption but before the liquidator purchases it. For instance, if one ETH is absorbed at $\$3000$ and its price declines to $\$2500$ before liquidation, the protocol incurs a loss.

\subsection{Borrow Collateral Factor Update Methodology}

To improve user borrowing capacity while effectively mitigating the risks of liquidation during periods of market volatility, we present a comprehensive methodology for calculating the Borrowing Collateral Factor (BCF).

\textbf{Methodology}:
The determination of the CF is grounded in a comprehensive analysis that considers several critical factors, including:

\begin{itemize}
    \item \textbf{Volatility}: Assets with higher historical volatility are associated with greater price fluctuations, which can increase the likelihood of liquidation events. Higher historical volatility of an asset is indicative of greater risk, which may necessitate a lower Borrowing Collateral Factor (BCF). This is due to the increased likelihood of significant price fluctuations that could lead to liquidation.
    \item \textbf{Liquidity}: Liquidity is another crucial factor influencing the BCF. Assets that exhibit higher liquidity can support elevated BCFs due to their ability to be sold more easily in the event of a default. The capacity to quickly convert collateral into cash reduces the risk to the protocol, allowing for a more favorable borrowing environment.
    \item \textbf{Market Depth}: Market depth refers to the volume of buy and sell orders at various price levels for a given asset. A deeper market mitigates the risk of slippage during liquidation events, allowing for the execution of larger trades without significantly impacting the asset's price. The ability to execute large trades without significantly affecting the asset's price is crucial for maintaining stability.
    \item \textbf{Historical Performance}: The historical performance of an asset, including its default rates and price stability, provides valuable insights into its risk profile. An asset that has demonstrated stable price behavior and low default rates over time can justify a higher BCF. By analyzing past performance metrics, we can better assess the likelihood of future stability and the associated risks, leading to more informed BCF determinations.
\end{itemize}

\subsection{Liquidation Collateral Factor Update Methodology}
The Liquidation Collateral Factor (LCF) is a critical parameter within the Compound protocol, defining the ratio at which a user's loan becomes subject to liquidation. This section outlines a comprehensive methodology for determining the appropriate LCF adjustment based on various factors.

\textbf{Evaluation of Large Supplier Positions: }
To assess the feasibility of increasing an asset's LCF, the positions of the largest suppliers within the protocol are examined. This evaluation involves examining the health factors associated with these positions, which reflect the collateralization status of the largest accounts.

If the positions of the largest suppliers are well-collateralized and exhibit adequate health factors, it may indicate that the protocol can accommodate a higher LCF without significantly increasing the risk of liquidation.

\textbf{Analysis of DEX Liquidity Levels:}
Next, we evaluate the liquidity levels of the asset on decentralized exchanges (DEXs). Adequate DEX liquidity is essential for effectively managing the offloading of positions in the event of a liquidation, particularly if an increase in the LCF leads to heightened borrowing activity.

By analyzing DEX liquidity levels, we can ensure that DEX liquidity can absorb the potential influx of sell orders during a liquidation event is crucial for mitigating the risk of significant price slippage.

\textbf{Consideration of Demand and Capital Efficiency:}
If DEX liquidity is sufficient and there is high demand for the asset within the protocol, increasing the LCF can enhance capital efficiency. A higher LCF allows users to borrow more against their collateral, optimizing asset utilization. Conversely, if liquidity is insufficient or demand is low, decreasing the LCF may be necessary to reduce liquidation risk.

Based on the evaluations of supplier positions, DEX liquidity levels, and asset demand, the LCF can be adjusted accordingly.

\begin{itemize}
    \item Increasing the LCF: If conditions are favorable—characterized by strong supplier health factors and adequate liquidity—an increase in the LCF is warranted to optimize capital efficiency and borrowing capacity.
    \item Decreasing the LCF: Conversely, if the analysis reveals potential risks, such as low liquidity or weak demand, the LCF may be maintained at its current level or decreased to prioritize the stability and risk management of the protocol.
\end{itemize}

\section{Results}

\textbf{Liquidations at Risk (LaR) analysis}

\begin{figure}[]
     \centering
         \includegraphics[width=\textwidth, trim=0cm 0cm 0cm 1.5cm]{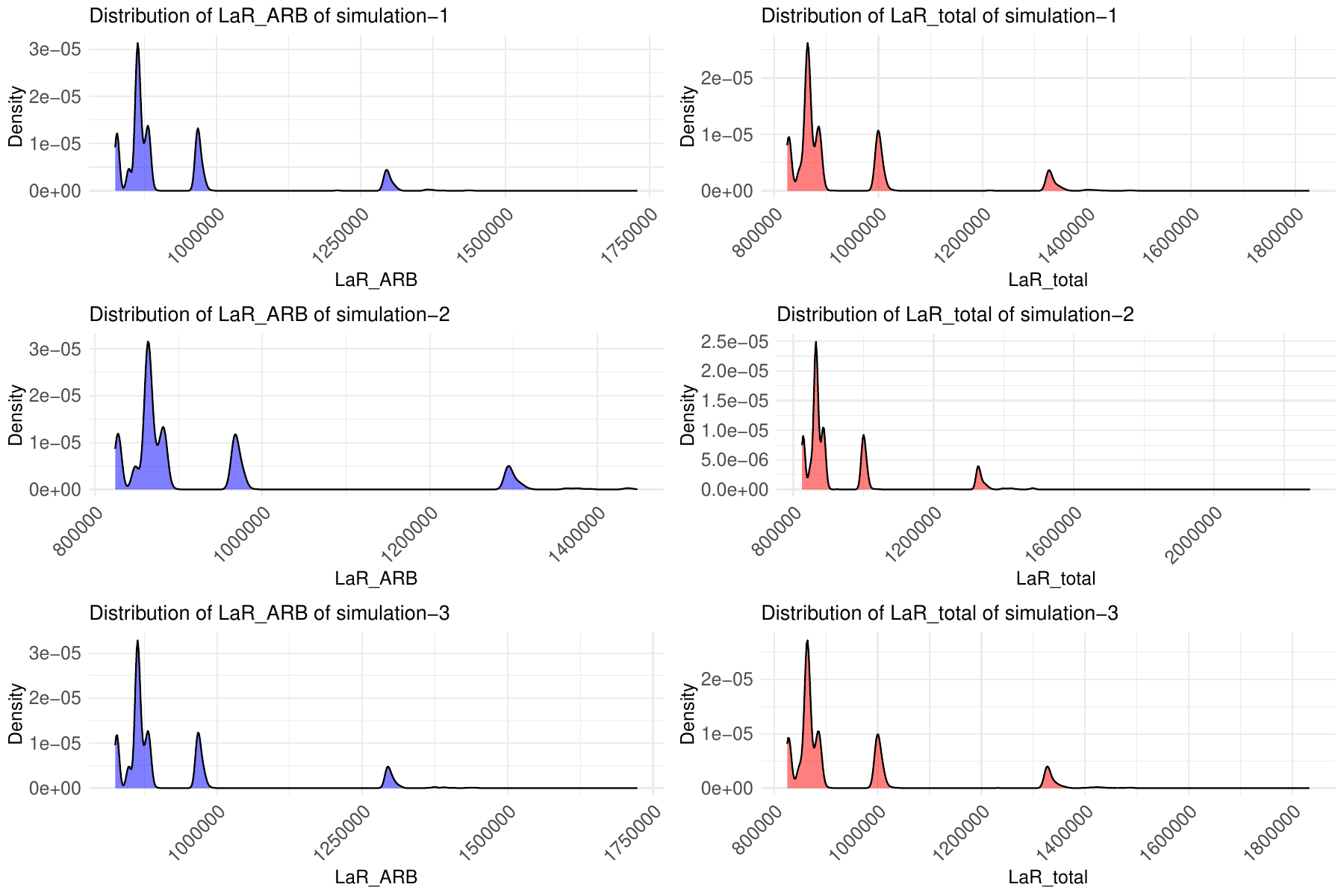}
         \caption{Liquidation at Risk (LaR) analysis.}
         \label{fig:lar_graphs}
\end{figure}

Figure \ref{fig:lar_graphs} depicts the probability density function of Liquidation at Risk (LaR) for Arbitrum (ARB) asset and overall (considering all assets combined). From Figure \ref{fig:lar_graphs}, it can be observed that each of the three simulation sets (total and asset-wise) employs the same distribution model. This denotes equal risk characteristics (nearly the same LaR distribution curve) across three sets, encompassing all extreme market conditions. In each simulation set, scenarios are generated randomly and independently. Such randomness helps prevent one scenario from having an undue effect on LaR estimates as a whole.


Every simulation set aims to cover all possible scenarios that can occur for purposes of liquidity analysis. By including extreme events, a plausible liquidity hazard evaluation can more easily be made.


The distribution function curve for estimated LaR values are more or less similar for each of the set of price trajectories. Thus, the estimator employed for calculating LaR is statistically consistent. It is to be further noticed that even if more such sets of price trajectories are created, it consistently produces accurate estimates. This consistent behavior is vital for ensuring that the LaR metric remains stable and dependable, even when applied to different sets of simulations.

Figure \ref{fig:lar_graphs} depicts the probability density function for LaR total and Arbitrum respectively.
The $LaR_{ARB}$ (blue plots) in all three simulations shows multiple peaks, meaning there are multiple scenarios of varying risks affecting the system. This could be due to changes in the market, types of collateral, or user actions that create different levels of liquidation risk. Each peak represents a group of similar risk levels. On the other hand, the $LaR_{total}$ (red plots) also has multiple peaks, but they differ from $LaR_{ARB}$, showing that the overall risk is influenced by many combined factors. This difference suggests that while Arbitrum has its own risk patterns, the total risk comes from a mix of several sources. 
$LaR_{GMX}$, $LaR_{WBTC}$, $LaR_{WETH}$: These metrics exhibit the same behavior across all three simulations. They are unimodal, showing a concentration around a single central value, which suggests a single dominant risk factor.



\section{Conclusion}
The economic audit of Compound V3 conducted by Chainrisk Labs, has yielded promising results, confirming that the protocol's design is well-aligned with its strategic objectives. Rigorous simulations and stress tests demonstrate that the protocol performs robustly under various scenarios.

Our findings demonstrate Compound V3’s resilience to scale and withstand high volatility across various collateral types. Even under extreme historical volatility conditions for assets like ETH, WBTC, ARB, and GMX the system remained significantly over-collateralized. Current liquidation incentives appear sufficient to manage risk. For collateral exhibiting super-linear slippage, more aggressive liquidation incentives and collateralization ratios may be necessary.

However, given the dynamic nature of user and market behavior, ongoing parameter evaluation and adjustment are crucial. Also, it is essential for Compound V3 to implement a continuous regime of regular audits, incorporate community feedback, and adapt to evolving market conditions to ensure ongoing stability and resilience. 

\section{Acknowledgement}
The authors would acknowledge the meaningful contribution of Varun Doshi, Rajdeep Sengupta, Vidhi Agarwal, Debajyoti Saha, Rohan Jha, and Anupam Shah in this article (audit report). They have been closely involved with their respective expertise while this article has been framed. The authors would like to extend their thanks to `allthecolors' for giving us constructive suggestions and comments in making our article exhaustive and more elaborate. 

\vspace{10mm}


\begin{thebibliography}{99}

\bibitem{chen2023web3}
Chen, Hongzhou and Duan, Haihan and Abdallah, Maha and Zhu, Yufeng and Wen, Yonggang and Saddik, Abdulmotaleb El and Cai, Wei.
\newblock Web3 Metaverse: State-of-the-art and vision.
\newblock \emph{ACM Transactions on Multimedia Computing, Communications and Applications}, 20(4):1--42, 2023.
\newblock Publisher: ACM New York, NY.

\bibitem{rutskiy2023dao}
Rutskiy, Vladislav and Muda, Iskandar and Joudar, Fadoua and Ilia, Filippov and Lyubaya, Svetlana and Kuzmina, Alexandra and Tsarev, Roman.
\newblock DAO Tokens: The Role for the Web 3.0 Industry and Pricing Factors.
\newblock In \emph{Computer Science On-line Conference}, pages 595--604, 2023.
\newblock Organization: Springer.

\bibitem{jensen2021introduction}
Jensen, Johannes Rude and von Wachter, Victor and Ross, Omri.
\newblock An introduction to decentralized finance (DeFi).
\newblock \emph{Complex Systems Informatics and Modeling Quarterly}, 26:46--54, 2021.

\bibitem{werner2022sok}
Werner, Sam and Perez, Daniel and Gudgeon, Lewis and Klages-Mundt, Ariah and Harz, Dominik and Knottenbelt, William.
\newblock Sok: Decentralized finance (DeFi).
\newblock In \emph{Proceedings of the 4th ACM Conference on Advances in Financial Technologies}, pages 30--46, 2022.

\bibitem{aramonte2021defi}
Aramonte, Sirio and Huang, Wenqian and Schrimpf, Andreas.
\newblock DeFi risks and the decentralisation illusion.
\newblock \emph{BIS Quarterly Review}, 6, 2021.
\newblock Publisher: Bank for International Settlements.

\bibitem{harrison2010introduction}
Harrison, Robert L.
\newblock Introduction to Monte Carlo simulation.
\newblock In \emph{AIP Conference Proceedings}, volume 1204, page 17, 2010.
\newblock Organization: NIH Public Access.

\bibitem{jackel2002monte}
Jäckel, Peter.
\newblock \emph{Monte Carlo methods in finance}, volume 5.
\newblock John Wiley \& Sons, 2002.

\bibitem{harvey2021defi}
Harvey, Campbell R and Ramachandran, Ashwin and Santoro, Joey.
\newblock \emph{DeFi and the Future of Finance}.
\newblock John Wiley \& Sons, 2021.

\bibitem{ugurlu2014modeling}
Ugurlu, Erginbay and Thalassinos, Eleftherios and Muratoglu, Yusuf.
\newblock Modeling volatility in the stock markets using GARCH models: European emerging economies and Turkey.
\newblock 2014. Publisher: Eleftherios Thalassinos.

\bibitem{guidolin2018autoregressive}
Guidolin, M and Pedio, M.
\newblock Autoregressive Moving Average (ARMA) models and their practical applications.
\newblock \emph{Essentials Time Series Financ. Applicat}, 41--76, 2018.

\bibitem{gauntletreport2020compound}
Gauntlet Research Report.
\newblock Market Risk Assessment.
\newblock February, 2020.

\end{thebibliography}



\end{document}